\documentclass[twocolumn,floatfix]{revtex4}
\pdfoutput=1
\usepackage{graphicx}
\usepackage{dcolumn}
\usepackage{longtable}
\usepackage{amsmath}
\usepackage{amssymb}
\usepackage{pdflscape}

\begin{document}

\title
{Proxy-SU(3) symmetry in heavy deformed nuclei}

\author
{Dennis Bonatsos$^1$, I. E. Assimakis$^1$, N. Minkov$^2$, Andriana Martinou$^1$, R. B. Cakirli$^3$, R. F. Casten$^{4,5}$, 
K. Blaum$^6$}

\affiliation
{$^1$Institute of Nuclear and Particle Physics, National Centre for Scientific Research 
``Demokritos'', GR-15310 Aghia Paraskevi, Attiki, Greece}

\affiliation
{$^2$Institute of Nuclear Research and Nuclear Energy, Bulgarian Academy of Sciences, 72 Tzarigrad Road, 1784 Sofia, Bulgaria}

\affiliation
{$^3$ Department of Physics, University of Istanbul, Istanbul, Turkey} 

\affiliation 
{$^4$ Wright Laboratory, Yale University, New Haven, Connecticut 06520, USA}

\affiliation
{$^5$ Facility for Rare Isotope Beams, 640 South Shaw Lane, Michigan State University, East Lansing, MI 48824 USA}

\affiliation
{$^6$ Max-Planck-Institut f\"{u}r Kernphysik, Saupfercheckweg 1, D-69117 Heidelberg, Germany}

\begin{abstract}

\parindent=0pt   
\textbf{Background:} Microscopic calculations of heavy nuclei face considerable difficulties due to the sizes of the matrices 
that need to be solved. Various approximation schemes have been invoked, for example by truncating the spaces, imposing seniority 
limits, or appealing to various symmetry schemes such as pseudo-SU(3). This paper proposes a new symmetry scheme also based on SU(3). This proxy-SU(3) can be applied to well-deformed nuclei,  is simple
to use, and can yield analytic predictions. 

\textbf{Purpose:} To present the new scheme and its microscopic motivation, and to test it using a Nilsson model calculation 
with the original shell model orbits and with the new proxy set. 

\textbf{Method:} We invoke an approximate, analytic, treatment of the Nilsson model, that allows the above vetting and yet 
is also transparent in understanding the approximations involved in the new proxy-SU(3).  

\textbf{Results:} It is found that the new scheme yields a Nilsson diagram for well-deformed nuclei that is very close to the 
original Nilsson diagram. The specific levels of approximation in the new scheme are also shown, for each major shell. 

\textbf{Conclusions:} The new proxy-SU(3) scheme is a good approximation to the full set of orbits in a major shell. Being able to replace a complex shell model calculation with
a symmetry-based description now opens up the possibility to
predict many properties of nuclei analytically and often in a parameter-free way.
The new scheme works best for heavier nuclei, precisely where full microscopic calculations are most challenged. Some cases in which the new scheme can be used, often analytically, to make specific predictions, are shown in a subsequent paper. 

\end{abstract}
 
\maketitle

\section{Introduction}

Microscopic approaches to the structure of atomic nuclei are becoming increasingly sophisticated and complex, and have made great strides, enabled by the rapid growth in the feasibility of computer intensive approaches using large bases and sophisticated interactions.  Nevertheless, realistic calculations of many observables in medium and heavy mass nuclei, or in exotic nuclei generally where there may be many valence nucleons, still (and for the foreseeable future) impose the need to invoke various simplifications, truncations, and approximations. There are many examples of such methods, ranging from straightforward limitations on the Hilbert space used, to, for example, seniority restrictions, or to symmetry-based approximation schemes such as pseudo-SU(3).  

Indeed, the present paper is inspired by the idea and success of pseudo-SU(3) but is based on a different substitution founded in the recognition that pairs of Nilsson orbits, $K[N n_z \Lambda]$, that are related by quantum numbers differing by 0[110], have high spatial overlap and identical angular momentum projection behavior. This leads to an approximate oscillator shell symmetry.  Of course, being based on an idealized symmetry, it is not at all a replacement for detailed microscopic shell model or ab initio calculations which are achieving more and more success, with great potential for the future.  However,  it does allow the possibility of very simple, analytic, parameter free predictions of certain nuclear properties, related especially to collective properties and nuclear shapes, that are robustly dependent on counting the number of nucleons (which determines the irreps of an applicable symmetry) interacting under a quadrupole interaction. It is applicable to deformed nuclei and provides the most advantages in heavier nuclei. 

In particular, the immediate purpose of this paper is to present a Nilsson calculation that demonstrates that this new scheme, called proxy-SU(3), leads to an excellent approximation to the actual  Nilsson diagram. We carry out this calculation 
in a transparent way that illuminates both the key ingredients in the new scheme and the level of approximation it entails. 

In the end, having vetted the approximate scheme, one can then exploit it (which we do in a subsequent paper \cite{second}), to make specific predictions about the behavior of deformed nuclei. 

\section{SU(3) and nuclear deformation: The motivation and nature of a new approximate SU(3)-based symmetry scheme}

The relation of SU(3) symmetry to nuclear deformation was discovered by J. P. Elliott \cite{Elliott1,Elliott2} in the sd shell nuclei, 
in which its microscopic origins have been demonstrated. The SU(3) symmetry also appears in the framework of the microscopic symplectic model \cite{Rosensteel}, which can be seen as a generalization of the Elliott SU(3) scheme to more than one nuclear shell. Since then 
the SU(3) symmetry has been used in a number of models, including the interacting boson model (IBM) \cite{IA}, the fermion dynamical symmetry model (FDSM) \cite{FDSM}, and the interacting vector boson model (IVBM) \cite{Georgieva}, especially in heavier nuclei, where the LS coupling scheme of the Elliott model breaks down \cite{Talmi}. 
It also forms the rationale for pseudo-SU(3) \cite{pseudo1,pseudo2,DW1,DW2,Ginocchio}, which we will discuss below. 
Finally, a quasi-SU(3) symmetry \cite{Zuker1,Zuker2}, based on the smallness of $\Delta j=1$ matrix elements,
leads to an approximate restoration of LS coupling in heavy nuclei.  

On the other hand, the Nilsson model \cite{Nilsson1,Nilsson2,RN}, despite its simplicity,  has been very successful in describing in detail many properties of heavy deformed nuclei. For large deformations, its wave functions reach an asymptotic limit, in which the number of oscillator quanta, $N$, the number of quanta 
along the cylindrical symmetry axis, $n_z$, and the projections of the orbital angular momentum, $\Lambda$, and of the spin, $\Sigma$, along the symmetry axis become good quantum numbers. They remain rather good even at intermediate deformation values \cite{Nilsson2}. As a consequence, Nilsson states 
for even nuclei are labelled by $K[N n_z \Lambda]$, where $K=\Lambda+\Sigma$ is the projection of the total angular momentum along the symmetry axis.  

As remarked by B. Mottelson \cite{Mottelson}, the asymptotic quantum numbers of the Nilsson 
model can be seen as a generalization of Elliott's SU(3), applicable to heavy deformed nuclei. Working along this line, we demonstrate in the present paper that a proxy-SU(3) symmetry of the Elliott type can be developed in heavy deformed nuclei. In order to achieve 
this, we take advantage of the large overlap of pairs of Nilsson orbits related by $\Delta K [\Delta N \Delta n_z \Delta \Lambda]=0[110]$.  The high overlaps of such pairs have already been shown, in the case of proton-neutron pairs in the rare earth region \cite{Cakirli,Karampagia},
 to play a key role in the onset and development of nuclear deformation.

In the present work we also take advantage of nucleon pairs with Nilsson quantum numbers differing by $\Delta K [\Delta N \Delta n_z \Delta \Lambda]=0[110]$, but in a different way. Instead of focusing attention on proton-neutron pairs, we use proton-proton 
pairs and neutron-neutron pairs.  This approach turns out to be successful in several ways, since it reveals a proxy-SU(3) symmetry in heavy deformed nuclei, which can be used either for making predictions of nuclear properties within the SU(3) symmetry using algebraic methods, or it might be useful as an approximation scheme for simplifying shell model calculations in heavy deformed nuclei away from closed shells, that are not yet accessible because of computational constraints.  


\begin{figure}[htb]

\includegraphics[width=100mm]{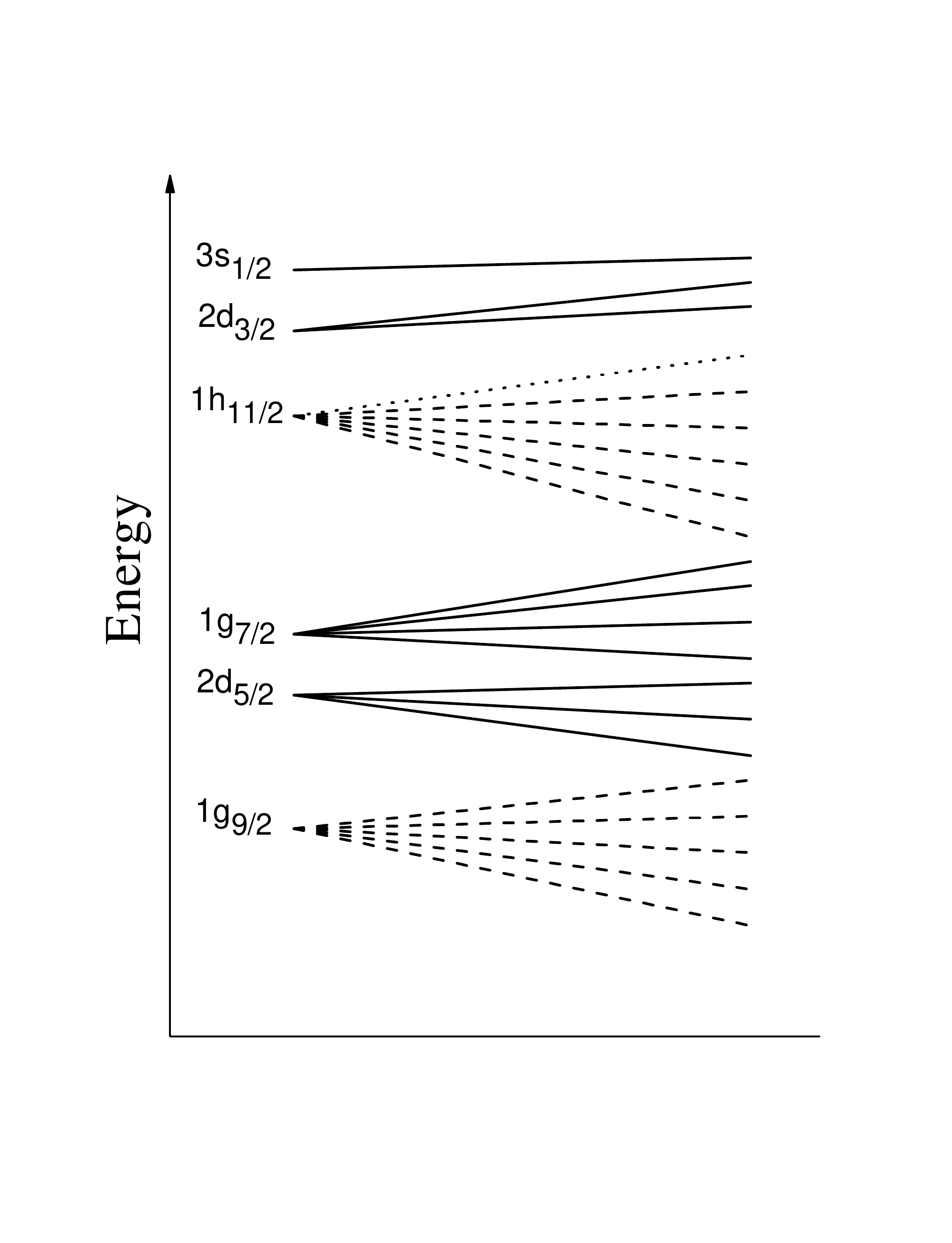}

\caption{Schematic representation of the 50--82 shell and the replacement leading to the proxy sdg shell. See Section II for further discussion.} 
 
\end{figure}

In order to see how this works, it is best to use a specific example, depicted in Fig.~1. We consider the 50-82 major shell and enumerate the following steps in the development of the new approximate scheme. 

1) The 50--82 major shell consists of the 3s$_{1/2}$, 2d$_{3/2}$, 2d$_{5/2}$, and 1g$_{7/2}$ orbitals (shown in Fig.~1 by solid lines), which are the pieces of the full sdg shell remaining after 
the spin-orbit force has lowered the 1g$_{9/2}$ orbitals (indicated by dashed lines) into the 28-50 nuclear shell. In addition, it contains the 1h$_{11/2}$ orbitals
(shown by dashed lines plus one dotted line), lowered 
into it from the pfh shell, also by the spin-orbit force. 

2) The 1g$_{9/2}$ orbital consists of the Nilsson orbitals 1/2[440], 3/2[431], 5/2[422], 7/2[413], 9/2[404].  Note that these are the 0[110] partners of the 
1h$_{11/2}$ Nilsson orbitals 1/2[550], 3/2[541], 5/2[532], 7/2[523], 9/2[514], in the same order.  A pair of these 0[110] partners shares exactly the same values of the quantum numbers corresponding to the projections of orbital angular momentum, 
spin, and total angular momentum. Thus the orbitals in such a pair are expected to exhibit identical behavior as far as 
properties related to angular momentum projection
are concerned. This has been corroborated by calculating overlaps of orbitals in Ref. \cite{Karampagia}. 

One can thus think of replacing all of the  1h$_{11/2}$ orbitals (the upper group of dashed lines in Fig.~1), except the 11/2[505]
orbital (the dotted line in Fig.~1) in the 50-82 shell by their 1g$_{9/2}$ counterparts (the lower group of dashed lines in Fig.~1)
and checking numerically the accuracy of this approximation, taking carefully into account that during this replacement the $N$ and $n_z$ quantum numbers have been changed by one unit each, while the parity has changed sign. 
These changes will obviously affect the selection rules of various relevant matrix elements, 
as well as the avoided crossings \cite{Cejnar} in the Nilsson diagrams, as we shall discuss in detail below.    

3) Note that the 1h$_{11/2}$ 11/2[505] orbit has been excluded here since it has no partner in the 1g$_{9/2}$ shell.  This is the sole orbit that has to be dropped in this approximation.  
It is important to recognize, however, that this orbit plays a very minor role in the evolution of structure in heavy nuclei, since it lies at the very top of the 50-82 shell in the Nilsson diagrams \cite{Nilsson1,Nilsson2}, and hence its influence is expected to be minimal. Moreover, it only comes into play in nuclei near the 82 shell closure that are not likely to be well-deformed in any case.
The same remark applies to analogous orbits in other shells such as the 13/2[606] orbit in the 82-126 shell.

4) After these two approximations have been made, we are left with a collection of orbitals which is exactly the same as the full sdg shell.
The sdg shell of the spherical harmonic oscillator is known to possess a U(15) symmetry, having an SU(3) subalgebra \cite{BK}. Therefore 
we can expect that some of the SU(3) features would appear within the approximate scheme. Of course one should bear in mind that in axially 
symmetric deformed nuclei the relevant symmetry is not spherical, but cylindrical \cite{Takahashi}. 
Therefore the relevant algebras are not U(N) Lie algebras, 
but more complicated versions of deformed algebras, in which, among the angular momentum operators, only the $L_z$ operator has the same physical content as the $L_z$ operator in the Nilsson model \cite{RD,ND,PVI,Lenis,Sugawara,Arima}. 

5) The same approach can be applied to the 28-50, 82-126, 126-184 shells, leading to approximate pf, pfh, sdgi shells, respectively, corresponding to 
U(10), U(21), U(28) algebras having SU(3) subalgebras (see \cite{BK} and references therein). 

6) An important consideration concerns the role and effect of level crossings in the Nilsson model. Orbits with different angular momentum and/or parity quantum numbers cannot interact, while 
in the case of identical angular momentum and parity, interactions and avoided crossings \cite{Cejnar} appear. As a result, one difference in the present approximate scheme in comparison to the original Nilsson model picture comes from the fact that in a nuclear shell individual particles in the intruder levels do not interact with those in the normal parity levels, while the proxies for
the intruder levels used in the present approximate scheme will interact with normal parity levels of the same angular momentum. These interactions are spurious and a consequence of our orbit substitution. Their effects need to be carefully assessed. For the current scheme to be useful, such interactions need to be small. We will show that this is the case in the next section. We also note that pair scattering can occur among both the shell model orbitals and those in our scheme.  The effect of this does not come into the treatment below of the Nilsson model for both situations, but will enter into practical calculations for actual observables. This will be addressed at the end and in a subsequent paper \cite{second}.            

\section{The Nilsson Hamiltonian for large deformations} 

The Nilsson single particle Hamiltonian \cite{Nilsson1,Nilsson2} is based on a harmonic oscillator with cylindrical symmetry 
supplemented with a spin-orbit term and an angular momentum squared term.  The Hamiltonian reads 
\begin{equation}
H=H_{osc} + v_{ls} \hbar \omega_0 ({\bf l} \cdot {\bf s}) + v_{ll} \hbar \omega_0 ({\bf l}^2 - \langle {\bf l}^2\rangle_N),
\label{Nil}\end{equation} 
where 
\begin{equation}
H_{osc}={ {\bf p}^2 \over 2M} +{1\over 2} M (\omega_z^2 z^2 + \omega^2_\perp (x^2+y^2))
\end{equation}
is the Hamiltonian of a harmonic oscillator with cylindrical symmetry. The quantity  
\begin{equation}
\langle {\bf l}^2\rangle_N= {1\over 2} N(N+3)
\end{equation}
is the average of the square of the angular momentum $\bf l$ within the $N$th oscillator shell,
$M$ is the nuclear mass, ${\bf s}$ is the spin, ${\bf p}$ is the momentum. 
The rotational frequencies $\omega_z$ and $\omega_\perp$ are related to the deformation parameter $\epsilon$ by
\begin{equation}
\omega_z = \omega_0 \left( 1 -{2\over 3} \epsilon \right), \qquad  \omega_\perp = \omega_0 \left( 1 +{1\over 3} \epsilon \right), 
\end{equation}
leading to 
\begin{equation}
\epsilon= {\omega_\perp - \omega_z \over \omega_0},
\end{equation}
with  $\epsilon > 0$ corresponding to prolate shapes and $\epsilon < 0$ corresponding to oblate shapes.
The standard values of the constants $v_{ls}$ and $v_{ll}$, determined from the available data on intrinsic nuclear spectra
\cite{BM}, are shown in Table I. In an alternative notation widely appearing in the literature \cite{Nilsson2}, the parameters 
$\kappa$ and $\mu$ are used, with $v_{ls}=-2\kappa$ and $v_{ll}=-\kappa \mu$. Their values are given in Table I. 

The eigenvalues of $H_{osc}$ are 
\begin{equation}\label{Hdiag}
E_{osc}= \hbar \omega_0 \left(N +{3\over 2} - {1\over 3} \epsilon (3n_z-N) \right). 
\end{equation}   

\begin{table}

\caption{Parameters $v_{ls}$ and $v_{ll}$  \cite{BM} used in the Nilsson Hamiltonian of Eq.~(1).
The corresponding values of the parameters $\kappa$ and $\mu$ of an alternative, widely used notation \cite{Nilsson2}, are also shown.}

\bigskip
\tabcolsep=8pt
\begin{tabular}{ r r r r r    }
                  
  region & $v_{ls}$  & $v_{ll}$ & $\kappa$ & $\mu$ \\

\hline
    
 $N,Z < 50$     & $-0.16$   & 0          & 0.08   & 0     \\
 $50 < Z < 82$  & $-0.127$  & $-0.0382$  & 0.0635 & 0.602 \\
 $82<N<126$     & $-0.127$  & $-0.0268$  & 0.0635 & 0.422 \\
 $82<Z<126$     & $-0.115$  & $-0.0375$  & 0.0575 & 0.652 \\
 $126<N$        & $-0.127$  & $-0.0206$  & 0.0635 & 0.324 \\
 
 \hline
\end{tabular}

\end{table}

Taking advantage of the cylindrical symmetry, and using the standard creation and annihilation operators $a^\dagger_x$, $a_x$, $a^\dagger_y$, $a_y$ for the quanta of the harmonic oscillator 
in the Cartesian coordinates $x$ and $y$, one can define creation and annihilation operators \cite{Nilsson2,MN}
\begin{multline}
R^+ ={1\over \sqrt{2}} (a_x^\dagger + i a_y^\dagger), \qquad R ={1\over \sqrt{2}} (a_x  - i a_y), \\
S^+ ={1\over \sqrt{2}} (a_x^\dagger - i a_y^\dagger), \qquad S ={1\over \sqrt{2}} (a_x +  i a_y),
\end{multline}
satisfying the commutation relations 
\begin{equation}
[R, R^\dagger]=[S,S^\dagger]=1, 
\end{equation}
thus going over to a $|n_z r s \Sigma \rangle $ basis, where $r$ is the number of quanta related to the harmonic oscillator
formed by $R^\dagger$ and $R$, and  $s$ is the number of quanta related to the harmonic oscillator
formed by $S^\dagger$ and $S$, for which 
\begin{equation}\label{rs}
n_\perp = r+s =N-n_z, \qquad \Lambda = r-s
\end{equation}
hold, where $n_\perp$ is the number of quanta perpendicular to the $z$-axis.  
It is then a straightforward task, described in detail in Ref. \cite{Nilsson2}, to calculate the matrix elements of the ${\bf l} \cdot {\bf s}$ and ${\bf l}^2$
operators in the new basis, the explicit results being given in Appendix I. 

Note that our approach here is to use the asymptotic wave functions (suitable for well-deformed nuclei) so that we can obtain 
analytic matrix elements for each term in the Hamiltonian. This enables us to isolate and explicitly exhibit the facets 
of our replacement scheme and to compare those with the traditional shell model level sequences. Our numerical solutions are 
therefore not identical to those of the usual Nilsson diagrams, although they are very close to them for $\epsilon>0.15$. 

\begin{table}

\caption{Nilsson model states in the $K[N n_z \Lambda]$ and $|n_z r s \Sigma \rangle$ notation,
in which $\Sigma=+1/2$ is represented by $+$ and $\Sigma=-1/2$ is represented by $-$. 
See Section III for further discussion.}

\bigskip
\tabcolsep=2pt
\begin{tabular}{ r r | r r | r r    }
                  
\hline 
$K[N n_z \Lambda]$ & $|n_z r s \Sigma \rangle$ & $K[N n_z \Lambda]$ & $|n_z r s \Sigma \rangle$ & $K[N n_z \Lambda]$ &
 $|n_z r s \Sigma \rangle$ \\
\hline
{\bf pf}       & {\bf pf}   &{\bf pfh}      &{\bf pfh}  &{\bf sdgi}     &{\bf sdgi} \\
1/2[301] & 021$-$ &1/2[501] &032$-$ &1/2[611] &132$-$ \\
1/2[321] & 210$-$ &1/2[521] &221$-$ &1/2[600] &033+ \\
3/2[312] & 120$-$ &3/2[512] &131$-$ &3/2[602] &042$-$ \\
1/2[310] & 111+ &1/2[510] &122+ &1/2[631] &321$-$ \\
3/2[301] & 021+ &3/2[501] &032+ &3/2[622] &231$-$ \\
5/2[303] & 030$-$ &5/2[503] &041$-$ &5/2[613] &141$-$ \\
1/2[330] & 300+ &1/2[541] &410$-$ &1/2[620] &222+ \\
3/2[321] & 210+ &3/2[532] &320$-$ &3/2[611] &132+ \\
5/2[312] & 120+ &5/2[523] &230$-$ &5/2[602] &042+ \\
7/2[303] & 030+ &7/2[514] &140$-$ &7/2[604] &051$-$ \\
         &      &1/2[530] &311+ &1/2[651] &510$-$ \\
{\bf sdg}      &{\bf sdg}   &3/2[521] &221+ &3/2[642] &420$-$ \\
1/2[400] & 022+ &5/2[512] &131+ &5/2[633] &330$-$ \\
1/2[411] & 121$-$ &7/2[503] &041+ &7/2[624] &240$-$ \\
3/2[402] & 031$-$ &9/2[505] &050$-$ &9/2[615] &150$-$ \\
1/2[420] & 211+ &1/2[550] &500+ &1/2[640] &411+ \\
3/2[411] & 121+ &3/2[541] &410+ &3/2[631] &321+ \\
5/2[402] & 031+ &5/2[532] &320+ &5/2[622] &231+ \\
1/2[431] & 310$-$ &7/2[523] &230+ &7/2[613] &141+ \\
3/2[422] & 220$-$ &9/2[514] &140+ &9/2[604] &051+ \\
5/2[413] & 130$-$ &11/2[505]&050+ &11/2[606]&060$-$ \\
7/2[404] & 040$-$ &         &     &1/2[660] &600+ \\
1/2[440] & 400+ & {\bf pfhj}    & {\bf pfhj}&3/2[651] &510+ \\
3/2[431] & 310+ &1/2[770] &700+ &5/2[642] &420+ \\
5/2[422] & 220+ &3/2[761] &610+ &7/2[633] &330+ \\
7/2[413] & 130+ &5/2[752] &520+ &9/2[624] &240+ \\
9/2[404] & 040+ &7/2[743] &430+ &11/2[615]&150+ \\
         &      &9/2[734] &340+ &13/2[606]&060+ \\
         &      &11/2[725]&250+ &         &     \\
         &      &13/2[716]&160+ &         &     \\
         &      &15/2[707]&070+ &         &     \\

 \hline
\end{tabular}

\end{table}

The correspondence between states in the $| n_z r s \Sigma \rangle$ basis and the standard Nilsson orbitals $K[ N n_z \Lambda]$
can be easily obtained using Eq. (\ref{rs}) and $K=\Lambda+\Sigma$, and is given in Table II. Results for the matrix elements  of ${\bf l} \cdot {\bf s}$ for the 50--82 and sdg shells are given in Table III and those for the matrix elements of ${\bf l}^2$ for the same shells are given in Table IV.  It should be remembered that the ${\bf l} \cdot {\bf s}$ and ${\bf l}^2$ terms are already relatively small perturbations of the oscillator potential for well-deformed nuclei, therefore the effects of the deformation on them can be neglected \cite{BM},
since they would correspond to second order corrections. As a result, the ${\bf l} \cdot {\bf s}$ and ${\bf l}^2$ matrix
elements appearing in Tables III and IV are (within the approximation used) independent of the deformation.  

The calculation of the energy eigenvalues of the full Hamiltonian becomes then a simple task of diagonalization of a matrix in which the diagonal matrix elements depend on the deformation, as given in Eq. (\ref{Hdiag}), while the non-diagonal matrix elements remain invariant, as stated in the last paragraph.  Therefore, the deformation enters formally only through the linear dependence of the diagonal matrix elements of the oscillator Hamiltonian on $\epsilon$.
In order for the proxy orbitals from the next lower shell to be brought to the energies of the intruder orbitals that they  replace,
their energies need to be uniformly pushed up by $1-2\epsilon/3$, as implied by Eq. (6), since both $N$ and $n_z$ have to be increased by one unit. Numerical results for $\epsilon =0.3$ for the 50--82 and sdg proton shells are given in Table V.
Nilsson-like diagrams involving either just the diagonal terms of the Hamiltonian, or 
obtained through the diagonalization of the full Hamiltonian, are plotted for the 50--82 and sdg proton shells in Fig. 2. Since the results have been obtained by using the asymptotic wave functions, they are expected to be reliable for large and moderate deformations \cite{Nilsson2}, but they should fail completely for  $\epsilon \leq 0.1$, where different approximate wave functions, providing different slopes of the energy levels as a function of $\epsilon$, are appropriate \cite{Nilsson2,BM}.

Results for the 28--50 and pf shells, the 82--126 and pfh shells, and the 126--184 and sdgi shells for the matrix elements of ${\bf l}^2$ and ${\bf l} \cdot {\bf s}$, as well as for those of the full Hamiltonian of Eq.~(1) are available at the end of the paper as Supplementary Material, which also contains figures similar to Figs. 2 and 3 below, for the other shells.  


\begin{table*}

\caption{${\bf l} \cdot {\bf s}$ matrix elements (in units of $\hbar \omega_0$) for Nilsson orbitals in the 50--82 shell (upper part) 
and in the full sdg shell (lower part), occurring after replacing the 1h$_{11/2}$ orbitals of the 50--82 shell by 
their 0[110] counterparts in the 1g$_{9/2}$ orbitals.  These matrix elements, and those in Tables IV and V, are calculated with the asymptotic Nilsson wave functions discussed in Section III. Therefore they are not precise eigenstates of the Nilsson Hamiltonian but go over to the latter in the limit of large deformation where the off-diagonal matrix elements are negligible compared to the diagonal ones. See subsection IV.A for further discussion. Note also that these are the matrix elements of the ${\bf l} \cdot {\bf s}$ operator alone. In the full Nilsson Hamiltonian of Eq. (1) they will be multiplied by the $v_{ls}$ coefficient  ($-0.127$) which will reduce their contributions to the Hamiltonian by a considerable factor.  Both matrices are symmetric, so only the diagonal and the upper half of each matrix are shown.  The new matrix elements appearing in the lower part of the table are shown in boldface. Otherwise, the upper and lower parts of the table are identical.  }

\bigskip
\tabcolsep=1pt
\begin{tabular}{ r r r r r r r r r r r | r r r r r r  }
                  &
${1\over 2}$[400] & ${1\over 2}$[411] & ${3\over 2}$[402] & ${1\over 2}$[420] & ${3\over 2}$[411] & ${5\over 2}$[402] & ${1\over 2}$[431] & 
${3\over 2}$[422] & ${5\over 2}$[413] & ${7\over 2}$[404] & ${1\over 2}$[550] & ${3\over 2}$[541] & ${5\over 2}$[532] & ${7\over 2}$[523] & 
${9\over 2}$[514] & ${11\over 2}$[505] \\

\hline

 1/2[400] & 0  & 1  & 0  & 0  & 0     & 0     & 0     & 0     & 0     & 0     & 0     & 0     & 0     & 0  & 0  & 0     \\
 1/2[411] &    & $-0.5$  & 0  & $-1.414$  & 0     & 0     & 0     & 0     & 0     & 0     & 0     & 0     & 0  & 0  & 0  & 0  \\
 3/2[402] &   &   & $-1$  & 0  & $-1.225$ & 0     & 0     & 0     & 0     & 0     & 0     & 0     & 0     & 0  & 0  & 0     \\
 1/2[420] &   &   &   & 0  & 0     & 0     & 1.225     & 0     & 0     & 0     & 0     & 0     & 0     & 0  & 0  & 0  \\
 3/2[411] &  &     &  &      & 0.5    & 0    & 0     & 1    & 0    & 0    & 0    & 0    & 0    & 0     & 0     & 0 \\
 5/2[402] &  &      &      &     &     & 1    & 0    & 0    & 0.707    & 0    & 0    & 0    & 0    & 0     & 0     & 0    \\
 1/2[431] &  &      &    &  &     &     & $-0.5$    & 0    & 0    & 0    & 0    & 0    & 0    & 0     & 0     & 0    \\
 3/2[422] &      &      &      &      &     &     &    & $-1$ & 0    & 0    & 0    & 0    & 0    & 0     & 0     & 0    \\
 5/2[413] &  &      &      &      &     &  &     &     & $-1.5$ & 0    & 0    & 0    & 0    & 0     & 0     & 0    \\
 7/2[404] &  &      &     &      &     &     &     &     &     & $-2$    & 0    & 0    & 0    & 0     & 0     & 0    \\
 
 \hline
 
 1/2[550] &  &      &      &      &     &     &    &     &     &     & 0    & 0    & 0    & 0     & 0     & 0    \\
 3/2[541] &  &      &      &      &    &     &     &     &     &     &     & 0.5    & 0    & 0     & 0     & 0    \\
 5/2[532] &  &      &      &      &     &     &     &     &     &     &     &     & 1    & 0     & 0     & 0    \\
 7/2[523] &   &   &   &   &      &      &      &      &     &      &      &      &      & 1.5  & 0  & 0     \\
 9/2[514] &   &   &   &   &      &      &      &      &      &      &      &      &      &   & 2  & 0     \\
11/2[505] &  &      &      &      &     &     &     &     &     &     &     &     &     &      &      & 2.5    \\
 
 \hline
\end{tabular}

\vskip 0.3cm

\begin{tabular}{ r r r r r r r r r r r | r r r r r   }
                  &
${1\over 2}$[400] & ${1\over 2}$[411] & ${3\over 2}$[402] & ${1\over 2}$[420] & ${3\over 2}$[411] & ${5\over 2}$[402] & ${1\over 2}$[431] & 
${3\over 2}$[422] & ${5\over 2}$[413] & ${7\over 2}$[404] & ${1\over 2}$[440] & ${3\over 2}$[431] & ${5\over 2}$[422] & ${7\over 2}$[413] & 
${9\over 2}$[404] \\

\hline

 1/2[400] & 0  & 1  & 0  & 0  & 0     & 0     & 0     & 0     & 0     & 0     & 0     & 0     & 0     & 0  & 0    \\
 1/2[411] &   & $-0.5$  & 0  & $-1.414$  & 0     & 0     & 0     & 0     & 0     & 0     & 0     & 0     & 0     & 0  & 0     \\
 3/2[402] &   &   & $-1$  & 0  & $-1.225$ & 0     & 0     & 0     & 0     & 0     & 0     & 0     & 0     & 0  & 0     \\
 1/2[420] &   &   &   & 0  & 0     & 0     & 1.225     & 0     & 0     & 0     & 0     & 0     & 0     & 0  & 0      \\
 3/2[411] &  &      &  &      & 0.5    & 0    & 0     & 1    & 0    & 0    & 0    & 0    & 0    & 0     & 0       \\
 5/2[402] &      &      &      &      &     & 1    & 0    & 0    & 0.707    & 0    & 0    & 0    & 0    & 0     & 0       \\
 1/2[431] & &      &      &  &     &   & $-0.5$    & 0    & 0    & 0  & ${\bf -1.414}$ & 0   & 0    & 0     & 0       \\
 3/2[422] &      &      &      &      &     &     &     & $-1$ & 0    & 0    & 0 & ${\bf -1.732}$ & 0   & 0     & 0       \\
 5/2[413] &      &      &      &      &     &  &    &    & $-1.5$ & 0    & 0    & 0   & ${\bf -1.732}$ & 0    & 0       \\
 7/2[404] &      &      &      &      &     &     &     &     &     & $-2$    & 0    & 0    & 0   & ${\bf -1.414}$ & 0     \\
 
 \hline
 
 1/2[440] &      &      &     &      &    &   & &     &     &     & 0    & 0    & 0    & 0     & 0       \\
 3/2[431] &  &      &     &      &     &     &     &  &     &     &     & 0.5    & 0    & 0     & 0       \\
 5/2[422] &  &      &      &      &     &     &     &     &     &     &     &     & 1    & 0     & 0        \\
 7/2[413] &   &   &   &   &      &      &      &      &      &     &      &      &      & 1.5  & 0    \\
 9/2[404] &   &   &   &   &      &      &      &      &      &      &      &      &      &   & 2     \\

 \hline
\end{tabular}

\end{table*}



\begin{table*}

\caption{Same as Table III, but for the ${\bf l}^2$ matrix elements for Nilsson orbitals in the 50--82 shell (upper part) 
and in the sdg shell (lower part). Note that these are the matrix elements of the ${\bf l}^2$ operator alone.  In the full Nilsson Hamiltonian of Eq. (1) they will be multiplied by the $v_{ll}$ coefficient ($-0.0382$) which will reduce their contributions to the Hamiltonian by a considerable factor.
See subsection IV.A for further discussion.}

\bigskip
\tabcolsep=1pt
\begin{tabular}{ r r r r r r r r r r r | r r r r r r  }
                  &
${1\over 2}$[400] & ${1\over 2}$[411] & ${3\over 2}$[402] & ${1\over 2}$[420] & ${3\over 2}$[411] & ${5\over 2}$[402] & ${1\over 2}$[431] & 
${3\over 2}$[422] & ${5\over 2}$[413] & ${7\over 2}$[404] & ${1\over 2}$[550] & ${3\over 2}$[541] & ${5\over 2}$[532] & ${7\over 2}$[523] & 
${9\over 2}$[514] & ${11\over 2}$[505] \\

\hline

 1/2[400] & 4  & 0  & 0  & $-5.657$ & 0     & 0     & 0     & 0     & 0     & 0     & 0     & 0     & 0     & 0  & 0  & 0     \\
 1/2[411] &   & 12  & 0  & 0  & 0     & 0     & $-6.928$ & 0     & 0     & 0     & 0     & 0     & 0     & 0  & 0  & 0     \\
 3/2[402] &   &   & 8   & 0  & 0 & 0     & 0     & $-4.899$    & 0     & 0     & 0     & 0     & 0     & 0  & 0  & 0     \\
 1/2[420] &  &   &   & 14  & 0     & 0     & 0     & 0     & 0     & 0     & 0     & 0     & 0     & 0  & 0  & 0     \\
 3/2[411] &     &      &  &      & 12    & 0    & 0     & 0    & 0    & 0    & 0    & 0    & 0    & 0     & 0     & 0    \\
 5/2[402] &     &     &    &    &   & 8    & 0    & 0    & 0    & 0    & 0    & 0    & 0    & 0     & 0     & 0    \\
 1/2[431] &    &  &      &  &     &    & 14    & 0    & 0    & 0    & 0    & 0    & 0    & 0     & 0     & 0    \\
 3/2[422] &   &  &   &      &     &    &     & 18 & 0    & 0    & 0    & 0    & 0    & 0     & 0     & 0    \\
 5/2[413] &      &      &  &      &     &  &     &    & 20 & 0    & 0    & 0    & 0    & 0     & 0     & 0    \\
 7/2[404] &      &      &      &      &     &     &     &     &     & 20    & 0    & 0    & 0    & 0     & 0     & 0    \\
 
 \hline
 
 1/2[550] &      &      &      &      &     &     &     &     &     &     & 10    & 0    & 0    & 0     & 0     & 0    \\
 3/2[541] &      &      &      &      &     &     &     &     &     &     &     & 18    & 0    & 0     & 0     & 0    \\
 5/2[532] &      &      &      &      &     &     &     &     &     &     &     &   & 24    & 0     & 0     & 0    \\
 7/2[523] &   &   &   &   &      &      &      &      &      &      &      &      &      & 28  & 0  & 0     \\
 9/2[514] &   &   &   &   &      &      &      &      &      &      &      &      &      &   & 30  & 0     \\
11/2[505] &      &      &      &      &     &     &     &     &     &     &     &     &     &      &      & 30    \\
 
 \hline
\end{tabular}

\vskip 0.3cm

\begin{tabular}{ r r r r r r r r r r r | r r r r r   }
                  &
${1\over 2}$[400] & ${1\over 2}$[411] & ${3\over 2}$[402] & ${1\over 2}$[420] & ${3\over 2}$[411] & ${5\over 2}$[402] & ${1\over 2}$[431] & 
${3\over 2}$[422] & ${5\over 2}$[413] & ${7\over 2}$[404] & ${1\over 2}$[440] & ${3\over 2}$[431] & ${5\over 2}$[422] & ${7\over 2}$[413] & 
${9\over 2}$[404] \\

\hline

 1/2[400] & 4  & 0  & 0  & $-5.657$ & 0     & 0     & 0     & 0     & 0     & 0     & 0     & 0     & 0     & 0  & 0       \\
 1/2[411] &   & 12  & 0  & 0  & 0     & 0     & $-6.928$ & 0     & 0     & 0     & 0     & 0     & 0     & 0  & 0     \\
 3/2[402] &  &   & 8   & 0  & 0 & 0     & 0     & $-4.899$    & 0     & 0     & 0     & 0     & 0     & 0  & 0       \\
 1/2[420] &  &   &   & 14  & 0     & 0     & 0     & 0     & 0     & 0     & ${\bf -6.928}$     & 0     & 0     & 0  & 0       \\
 3/2[411] &      &      &  &      & 12    & 0    & 0     & 0    & 0    & 0    & 0    & ${\bf -6.928}$    & 0    & 0     & 0        \\
 5/2[402] &      &      &    &      &     & 8    & 0    & 0    & 0    & 0    & 0    & 0    & ${\bf -4.899}$    & 0     & 0         \\
 1/2[431] &     &   &      &  &     &     & 14    & 0    & 0    & 0    & 0    & 0    & 0    & 0     & 0        \\
 3/2[422] &      &  &      &      &     &     &     & 18 & 0    & 0    & 0    & 0    & 0    & 0     & 0        \\
 5/2[413] &      &      &  &      &     &  &     &     & 20 & 0    & 0    & 0    & 0    & 0     & 0         \\
 7/2[404] &      &      &      &      &     &     &     &     &     & 20    & 0    & 0    & 0    & 0     & 0        \\
 
 \hline
 
 1/2[440] &      &      &      &  &     &     &     &    &     &     & {\bf 8}    & 0    & 0    & 0     & 0         \\
 3/2[431] &      &      &      &      &     &     &     &     &     &     &     & {\bf 14}    & 0    & 0     & 0         \\
 5/2[422] &      &      &      &      &     & &     &     &     &     &     &     & {\bf 18}    & 0     & 0         \\
 7/2[413] &   &   &   &   &      &      &      &      &      &      &      &      &      & {\bf 20}  & 0       \\
 9/2[404] &   &   &   &   &      &      &      &      &      &      &      &      &      &   & {\bf 20}      \\

 \hline
\end{tabular}

\end{table*}



\begin{table*}

\caption{Same as Table III, but for the $H$ matrix elements with $\epsilon=0.3$ for Nilsson orbitals in the 50--82 proton shell (upper part)
and in the sdg proton shell (lower part). Note again that these calculations use the asymptotic wave functions defined in Section III. The deformation comes in only through the linear dependence on epsilon of the diagonal matrix elements of 
$H_{osc}$. 
See subsection IV.A for further discussion.}

\bigskip

\begin{tabular}{ r r r r r r r r r r r | r r r r r r  }
                  &
${1\over 2}$[400] & ${1\over 2}$[411] & ${3\over 2}$[402] & ${1\over 2}$[420] & ${3\over 2}$[411] & ${5\over 2}$[402] & ${1\over 2}$[431] & 
${3\over 2}$[422] & ${5\over 2}$[413] & ${7\over 2}$[404] & ${1\over 2}$[550] & ${3\over 2}$[541] & ${5\over 2}$[532] & ${7\over 2}$[523] & 
${9\over 2}$[514] & ${11\over 2}$[505] \\

\hline

 1/2[400] & 6.28  & $-0.13$  & 0  & 0.22  & 0     & 0     & 0     & 0     & 0     & 0     & 0     & 0     & 0     & 0  & 0  & 0     \\
 1/2[411] &  & 5.74  & 0  & 0.18  & 0     & 0     & 0.27  & 0     & 0     & 0     & 0     & 0     & 0     & 0  & 0  & 0     \\
 3/2[402] &   &   & 6.26  & 0  & 0.16 & 0     & 0     & 0.19     & 0     & 0     & 0     & 0     & 0     & 0  & 0  & 0     \\
 1/2[420] &   &   &  & 5.30  & 0     & 0     & $-0.16$     & 0     & 0     & 0     & 0     & 0     & 0     & 0  & 0  & 0     \\
 3/2[411] &     &     &  &      & 5.61    & 0    & 0     & $-0.13$    & 0    & 0    & 0    & 0    & 0    & 0     & 0     & 0    \\
 5/2[402] &      &      &      &    &  & 6.00    & 0    & 0    & $-0.09$    & 0    & 0    & 0    & 0    & 0     & 0     & 0    \\
 1/2[431] &      &      &      & &     &    & 5.06    & 0    & 0    & 0    & 0    & 0    & 0    & 0     & 0     & 0    \\
 3/2[422] &   &      &     &      &    &     &   & 5.27 & 0    & 0    & 0    & 0    & 0    & 0     & 0     & 0    \\
 5/2[413] &    &    &    &   &  & &    &    & 5.56 & 0    & 0    & 0    & 0    & 0     & 0     & 0    \\
 7/2[404] &   &   &     &      &     &     &     &     &   & 5.93    & 0    & 0    & 0    & 0     & 0     & 0    \\
 
 \hline
 
 1/2[550] &    &   &    &     &   &   &   &   &   &    & 5.88   & 0    & 0    & 0     & 0     & 0    \\
 3/2[541] &    &    &   &    &   &   &   &   &  &  &  & 5.81    & 0    & 0     & 0     & 0    \\
 5/2[532] &  &    &  &  &  &   &  &  &  &  &   &  & 5.82    & 0     & 0     & 0    \\
 7/2[523] & & & &  &     &      &      &      &      &      &      &      &      & 5.90  & 0  & 0     \\
 9/2[514] &   &   &  &   &      &      &      &      &      &      &      &      &      &  & 6.06  & 0     \\
11/2[505] &      &      &      &     &  &   &   &   &   &  &   &    &   &     &   & 6.30    \\
 
 \hline
\end{tabular}

\vskip 0.3cm

\begin{tabular}{ r r r r r r r r r r r | r r r r r   }
                  &
${1\over 2}$[400] & ${1\over 2}$[411] & ${3\over 2}$[402] & ${1\over 2}$[420] & ${3\over 2}$[411] & ${5\over 2}$[402] & ${1\over 2}$[431] & 
${3\over 2}$[422] & ${5\over 2}$[413] & ${7\over 2}$[404] & ${1\over 2}$[440] & ${3\over 2}$[431] & ${5\over 2}$[422] & ${7\over 2}$[413] & 
${9\over 2}$[404] \\

\hline

 1/2[400] & 6.28  & $-0.13$  & 0  & 0.22  & 0     & 0     & 0     & 0     & 0     & 0     & 0     & 0     & 0     & 0  & 0      \\
 1/2[411] &   & 5.74  & 0  & 0.18  & 0     & 0     & 0.27  & 0     & 0     & 0     & 0     & 0     & 0     & 0  & 0       \\
 3/2[402] &  &  & 6.26  & 0  & 0.16 & 0     & 0     & 0.19     & 0     & 0     & 0     & 0     & 0     & 0  & 0     \\
 1/2[420] &   &  &   & 5.30  & 0 & 0 & $-0.16$     & 0     & 0     & 0     & {\bf 0.27}     & 0     & 0     & 0  & 0      \\
 3/2[411] &   &   & &     & 5.61    & 0    & 0     & $-0.13$    & 0    & 0    & 0    & {\bf 0.27}    & 0    & 0     & 0       \\
 5/2[402] &    &   &   &    &    & 6.00    & 0    & 0    & $-0.09$    & 0    & 0    & 0    & {\bf 0.19}    & 0     & 0       \\
 1/2[431] &   &     &      &  &    &    & 5.06    & 0    & 0    & 0    & {\bf 0.18}    & 0    & 0    & 0     & 0       \\
 3/2[422] &      &      &    &    &   &    &   & 5.27 & 0    & 0    & 0    & {\bf 0.22}    & 0    & 0     & 0       \\
 5/2[413] &  &    &    &    &   &  &    &   & 5.56 & 0    & 0    & 0    & {\bf 0.22}    & 0     & 0       \\
 7/2[404] &   &   &    &  &   &  &   &   &   & 5.93    & 0    & 0    & 0    & {\bf 0.18}     & 0       \\
 
 \hline
 
 1/2[440] &  &   &    &   &    &   &   &   &   &  & {\bf 5.73}   & 0    & 0    & 0     & 0        \\
 3/2[431] &  &  &   &   &   &    &  &    &    &    &   & {\bf 5.74}    & 0    & 0     & 0       \\
 5/2[422] &   &     &     &     &    &     &     &   &     &    &     &   & {\bf 5.82}    & 0     & 0       \\
 7/2[413] &   &   &   &   &      &     &      &      &      &      &      &      &     & {\bf 5.98}  & 0      \\
 9/2[404] &   &   &   &   &      &      &      &      &      &      &      &      &      &   & {\bf 6.22}       \\

 \hline
\end{tabular}

\end{table*}


\section{Discussion} 

\subsection{The ${\bf l} \cdot {\bf s}$, ${\bf l}^2$, and $H$  matrix elements}

In the upper half of Table III the matrix elements of the spin-orbit term in the 50--82 shell are shown.
These are compared to the relevant matrix elements appearing in the full sdg shell, 
that is, the shell comprising the 3s, 2d, and 1g$_{7/2}$ positive parity orbits in the 50--82 shell
and the 1g$_{9/2}$ orbit from the next lower shell that we are using as a proxy for the 1h$_{11/2}$ orbit,
seen in the lower half of Table III.
The boldface entries in the lower half of Table III are the new values appearing in the case of the modified shell
occurring after replacing the 1h$_{11/2}$ levels of the 50--82 shell (the last 6 levels in the rows and columns of the upper half of Table III) by their 0[110] counterparts of the 1g$_{9/2}$ levels (the last 5 levels in the rows and columns of the lower half of 
Table III). Each half of the table is divided into four blocks by straight lines. 

We see that the upper part of Table III has one more column (the last one)  and one more row (the last one) than 
the lower part of Table III because the 11/2[505] level of the 50--82 shell has no 0[110] counterpart in the sdg shell.  The upper left blocks of the two parts of the table are obviously identical, since they refer to the same set of states.  The lower right blocks of the two parts of the table are also identical, since the 0[110] pairs possess the same orbital angular momentum and spin quantum numbers, taking also into account that the last level of 1h$_{11/2}$, 11/2[505], has no counterpart in 1g$_{9/2}$.   Finally, the upper right block in the top half of Table III is ``empty'', since all matrix elements vanish, while 
in the bottom half of Table III a few non-vanishing matrix elements (4 out of 50 in each block) appear. 
They occur because the 1g$_{9/2}$ orbitals have non-vanishing matrix elements with the same parity orbits of the 50--82 shell,
while the opposite parity 1h$_{11/2}$ orbitals do not. These spurious matrix elements are a consequence of the proxy scheme 
we are invoking here. We will see that they are few and have very small effects, since within the total Hamiltonian they get multiplied by the small values of the coefficient $v_{ls}$, given in Table I. 
These spurious non-vanishing matrix elements represent the ``damage'' made by the approximation imposed. In what follows, it will be clear that the spurious  
matrix elements are responsible for a modification of the single particle energies resulting from the diagonalization of the full Hamiltonian matrix, 
as well as the occurence of interactions connecting in each shell the 0[110] proxies of the intruder orbitals with the normal parity orbitals, resulting in additional avoided crossings in the Nilsson diagrams. 

Similar comments apply to all other shells starting with 28--50, shown as Supplementary Material at the end of the paper. The numbers of spurious matrix elements for each shell are summarized in Table VI.  

\begin{table}

\caption{Number of matrix elements of the operators  ${\bf l} \cdot {\bf s}$,  ${\bf l}^2$, and  $H$
in the pf, sdg, pfh, and sdgi shells, differing from the corresponding matrix elements in the 28--50, 50--82, 82--126,
and 126--184 shells. The total number of matrix elements for each operator is given in the last column.
See subsection IV.A for further discussion.} 

\bigskip
\tabcolsep=12pt
\begin{tabular}{ r r r r r  }
                  
  shell & ${\bf l} \cdot {\bf s}$ & ${\bf l}^2$ &   $H$  & total  \\

\hline
    
 pf     &  6  &  - &  6  & 100 \\
 sdg    &  8  & 11 & 19  & 225 \\
 pfh    &  10 & 14 & 24  & 441 \\
 sdgi   &  12 & 17 & 29  & 784 \\

 \hline
\end{tabular}

\end{table}

Qualitatively similar results are obtained in the case of the matrix elements of the ${\bf l}^2$ operator, 
shown for the 50--82 and sdg shells in the upper and lower halves of Table IV reprectively, and in the Supplementary Material at the end of the paper for the rest of the shells, 
but also a few differences appear. In particular, no ${\bf l}^2$ term is used in the 28--50 and pf shells.  Also, in all pairs of shells, the diagonal matrix elements appearing in the lower right block are slightly different, since they depend on $n_z$, as seen in Eq.~(\ref{l2diag}).  Table VI also contains a similar list for the number of these spurious matrix elements.

Finally, in order to get a feeling of the number and magnitude of  matrix elements of the full Hamiltonian affected by the approximation, we present results for the special case of $\epsilon =0.3$  
for the 50--82 and sdg proton shells in the upper and lower halves of Table V reprectively, and in the Supplementary Material at the end of the paper for the rest of the shells. 
In each shell, the appropriate $v_{ls}$ and $v_{ll}$ values taken from Table I are used.  In all cases the numerical values of the diagonal matrix elements are at least one order of magnitude larger than the numerical values of the non-diagonal matrix elements. Table VI includes the number of spurious matrix elements for the full Hamiltonian as well.

From this analysis we see that a very small percentage of matrix elements are affected by the approximations made in the present scheme. In the sdg shell, 8.4\% are spurious. In heavier shells, this number drops to 5.4\% in the pfh shell  and to 3.7\% in the sdgi shell and, in general, our model is expected to work best the heavier the nucleus. Therefore the example shown in this discussion, the sdg shell, is the one in which the ``damages'' caused by the approximation are the most severe. 


\begin{figure*}[htb]

\includegraphics[width=150mm]{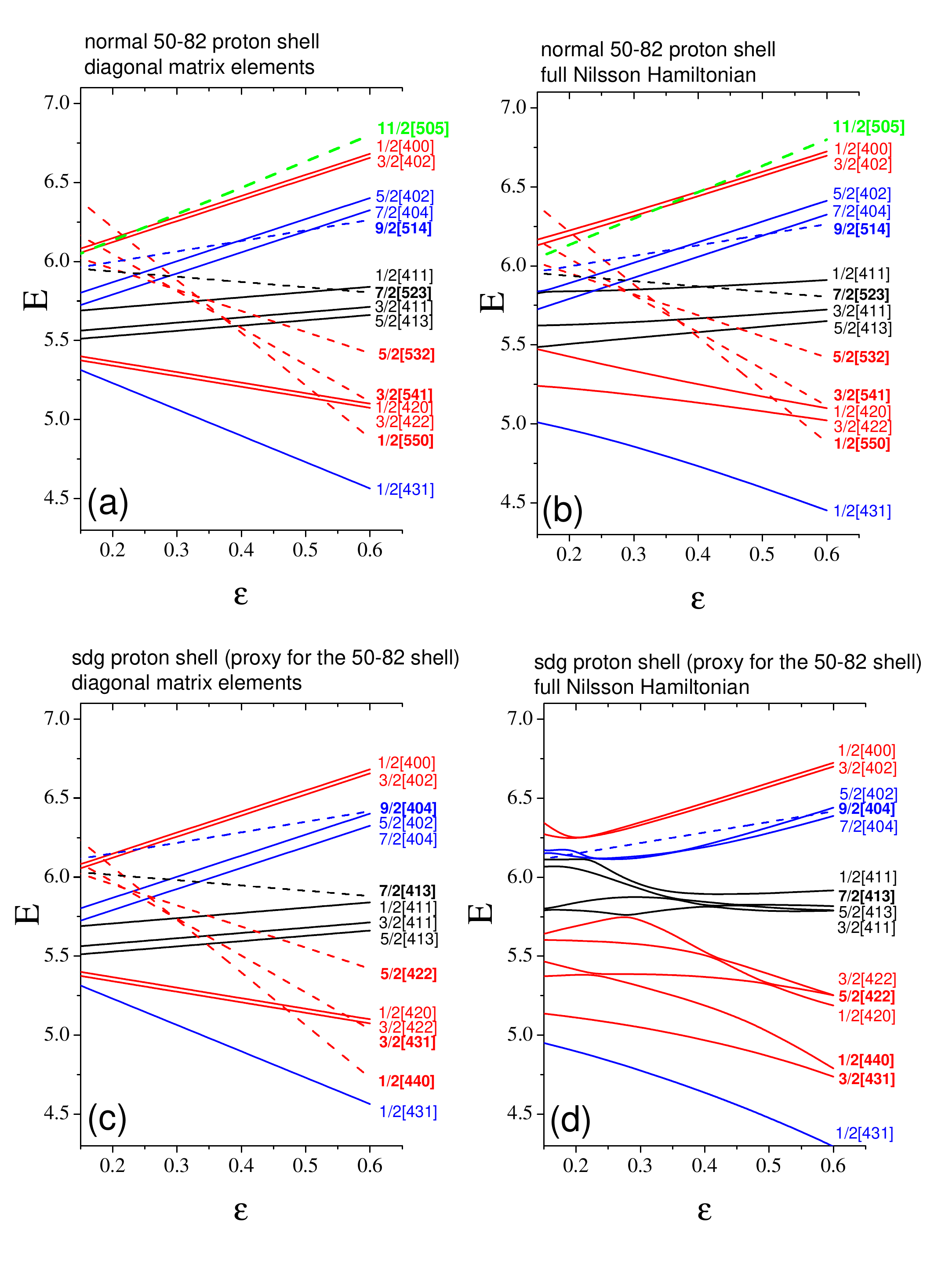}

\caption{(color online) Energies (in units of $\hbar \omega_0$) of the Nilsson Hamiltonian as functions of the deformation parameter $\epsilon$.
Note that, in these panels, the energies shown are not the usual solutions to the Nilsson Hamiltonian, but use matrix elements 
obtained analytically in the asymptotic deformed basis. Therefore they are not valid for small deformations and would not go to the usual degeneracies at $\epsilon=0$. They are, however, quite accurate for $\epsilon > 0.15$. The 
panels in this figure therefore also start at $\epsilon=0.15$.  The Nilsson parameters are taken from Table I. The 1h$_{11/2}$ orbitals in (a) and (b), as well as the 1g$_{9/2}$ orbitals in (c), are 
indicated by dashed lines. The 1h$_{11/2}$ orbital labels in (a) and (b), as well as the 1g$_{9/2}$ orbital labels in (c) and (d), appear
in boldface. Orbitals are grouped in color only to facilitate visualizing 
the patterns of orbital evolution. Note that the Nilsson labels at the right are always in the same order
as the energies of the orbitals as they appear at the right as well (largest deformation shown). 
Therefore, in some cases the order of the Nilsson orbitals changes slightly from panel to panel.
(a) Energies (diagonal matrix elements) for the 50--82 proton shell, including the contributions from both the $H_{osc}$ term and the small perturbations (the ${\bf l} \cdot {\bf s}$ and ${\bf l}^2$ terms). Therefore the Nilsson trajectories are straight 
and exhibit crossings. 
(b) Results of the full diagonalization for the 50--82 proton shell, in which the non-diagonal 
matrix elements are taken into account. Hence the Nilsson trajectories show the usual curvatures and avoided crossings. 
(c) Energies (diagonal matrix elements) for the sdg proton shell, resulting after the replacement of the 1h$_{11/2}$ orbitals of the 50--82 shell by their 0[110] 1g$_{9/2}$ counterparts. Therefore the Nilsson trajectories are straight and exhibit crossings.
(d) Results of the full diagonalization for the sdg proton shell. Hence the Nilsson trajectories again show the curvatures (enhanced relative to Fig. 2(b) by the mixing related to the spurious off-diagonal matrix elements of the 1g$_{9/2}$ orbit) and avoided crossings.  
See subsection IV.B for further discussion.
 } 
\end{figure*}


\begin{figure*}[htb]

\includegraphics[width=150mm]{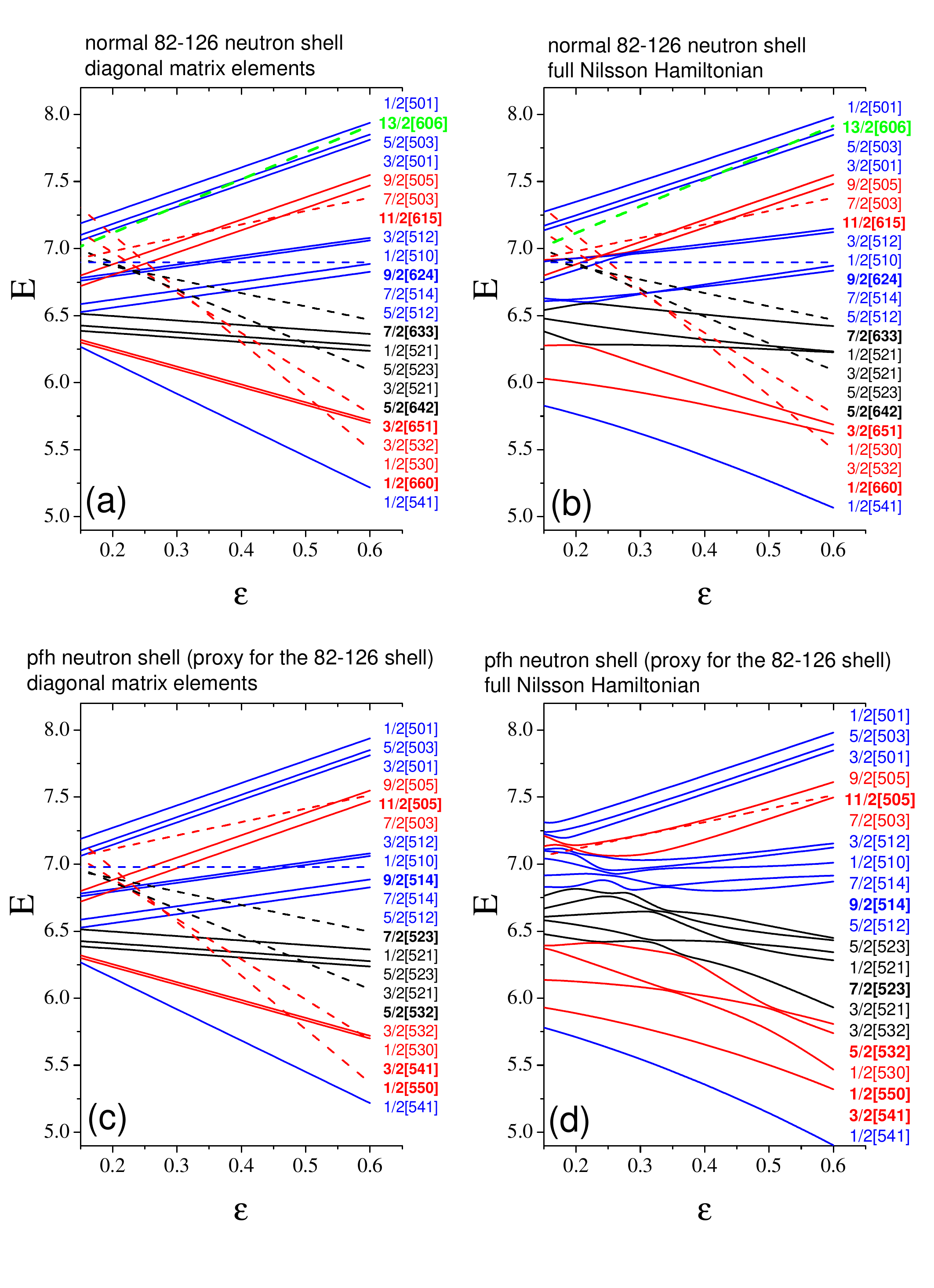}

\caption{(color online) Same as Fig.~2, but for the diagonal matrix elements (in units of $\hbar \omega_0$) of the Nilsson Hamiltonian for the 82-126 (a) and pfh (c) neutron shells compared 
to the results of the full diagonalization for the 82-126 (b) and pfh (d) neutron shells, as functions of the deformation parameter $\epsilon$. 
The Nilsson parameters are taken from Table I. The 1i$_{13/2}$ orbitals in (a) and (b), as well as the 1h$_{11/2}$ orbitals in (c), are 
indicated by dashed lines. The 1i$_{13/2}$ orbital labels in (a) and (b), as well as the 1h$_{11/2}$ orbital labels in (c) and (d), appear
in boldface. Orbitals are grouped in color only to facilitate visualizing 
the patterns of orbital evolution. Note that the Nilsson labels at the right are always in the same order
as the energies of the orbitals as they appear at the right as well (largest deformation shown). 
Therefore, in some cases the order of the Nilsson orbitals changes slightly from panel to panel.
 } 
\end{figure*}

\subsection{Nilsson orbit energies} 

Numerical results for the normal 50--82 proton shell are shown in Fig.~2(a),(b), while results for the sdg proton shell,
resulting after the replacement of the 1h$_{11/2}$ orbitals of the 50--82 shell by their 0[110] 1g$_{9/2}$ counterparts, are depicted in Fig.~2(c),(d). In Fig.~2(a),(c) the diagonal matrix elements are plotted,
including the contributions from both the $H_{osc}$ term and the small perturbations (the ${\bf l} \cdot {\bf s}$ and 
${\bf l}^2$ terms), while in Fig.~2(b),(d) the results of the diagonalization of the full Hamiltonian, in which the non-diagonal 
matrix elements are taken into account, are given. 

Note that these figures are not the same as the usual Nilsson diagrams since they use the approximate asymptotic basis, 
where analytic results can be obtained for all terms in the Hamiltonian. Thus these results are not valid (and not shown) 
for small deformations. In the well-deformed region, however, they closely approximate the usual diagonalizations of the 
Nilsson Hamiltonian in the $N l j \Omega$ basis \cite{Nilsson2}. In panels 2(a) and 2(c), where only the diagonal results 
are shown, the energies lie along straight lines without curvature. In 2(b) and 2(d) the full Hamiltonian with off-diagonal 
elements is used, resulting in the familiar curved trajectories and avoided crossings. 

To see the patterns in this figure, and in Fig.~3, more easily, we have color coded groups of orbits in the panels of Fig.~2 so that the locations of similar sequences of orbits can be identified at a glance. 

We first compare Fig.~2(a) with Fig.~2(c) (the two panels in the left hand side of the figure), i.e. the diagonal terms in the 50--82 and sdg proton shells.

In Fig.~2(a) the normal parity levels are shown as solid lines, while the six 
intruder 1h$_{11/2}$ orbitals are indicated by dashed lines.  

In Fig.~2(c) the normal parity levels, shown again as solid lines,  correspond to those in Fig.~2(a),
while the dashed lines indicate the five 1g$_{9/2}$ orbitals which have replaced the 1h$_{11/2}$ orbitals. 

We see that the five 1g$_{9/2}$ orbitals in Fig.~2(c) lie at positions very similar to those of their 0[110] partners 
in Fig.~2(a). For example, 1/2[550] of Fig.~2(a) and 1/2[440] of Fig.~2(c) lie at very similar positions. 

The high-lying 11/2[505] orbital of Fig.~2(a) has no analog in Fig.~2(c). 

We now compare Fig.~2(a) with Fig.~2(b) (the two top panels in the figure), i.e., we turn on the non-diagonal interactions in the 50--82 proton shell. 

The intruder 1h$_{11/2}$ orbitals, shown by dashed lines in both figures, are not affected, since there are no matrix elements 
connecting them to other orbitals, as is clear from the upper part of Table V. 

The normal parity orbitals, shown by solid lines in both figures, are changed by the non-diagonal matrix elements 
interconnecting them, as seen in the upper part of Table V. Since the non-diagonal matrix elements are at least one order of magnitude 
smaller than the diagonal matrix elements, the relative positions of the various lines are not affected much, although the mixing of states is clear from the small curvatures seen. 

We can now compare Fig.~2(c) with Fig.~2(d) (the two bottom panels in the figure), i.e., we turn on the non-diagonal interactions in the sdg proton shell. 

The 1g$_{9/2}$ orbitals, which have replaced the 1h$_{11/2}$ orbitals, do interact with the normal parity orbitals, a shortcoming of our approach, as seen in the lower part of Table V, with the exception of 9/2[404], which is still indicated by a dashed line in Fig.~2(d). However, as we shall see, the effects of this approximation are modest. As a result of the non-diagonal matrix elements, all lines in Fig.~2(d) (except 9/2[404]) are curved, forming, however, a figure quite similar to Fig.~2(c).

We finally compare Fig.~2(b) with Fig.~2(d) (the two panels in the right hand side of the figure), i.e., the final results for the 50--82 and sdg proton shells. 
The similarity of the two figures is clear. 

As noted above, the differences are due to the fact that the 1g$_{9/2}$ orbitals in Fig.~2(d) interact with the normal parity 
orbitals (see lower part of Table V) while the 1h$_{11/2}$ orbitals in Fig.~2(c) do not interact with the normal parity orbitals
(upper part of Table V). These additional interactions, which are spurious in the sense of arising from the introduction of the 1g$_{9/2}$ orbit in the approximate scheme we use, account for the  greater curvature in Fig.~2(d) than in Fig.~2(b).

Figure~3 shows similar results for the 82-126 and pdf neutron shells where the similarity of patterns 
is even more apparent than in Fig. 2.  
Similar plots can be made for other shells, being provided as Supplementary Material at the end 
of the paper.

In the case of the 28--50 and pf shells, non-diagonal matrix elements are contributed only by the ${\bf l} \cdot {\bf s}$
term, since no ${\bf l}^2$ term is used in these shells, resulting in very similar diagrams, since only 6\% of the matrix elements are affected by the approximation (see Table VI).   

In the cases of the 82--126 and pfh shells, as well as of the 126--184 and sdgi shells, the resulting diagrams are also very similar (as noted above for the first of these shells for Fig. 3), since the percentage of matrix elements affected by the approximation is very low, being 5.4\% and 3.7\% respectively (see Table VI). 

It should be noticed that the example of the 50--82 and sdg shells, shown in Fig.~2, is therefore {\sl the worst} possible one, 
since the percentage of the matrix elements affected by the approximation is highest, 8.4\% (again, see Table VI). Indeed, as noted above, the agreement in Fig.~3 between the exact 
and the approximate cases is greatly improved compared to Fig.~2.  

In all shells one sees that the changes inflicted on the Nilsson diagrams by the replacement of the intruder parity orbitals with their 0[110] counterparts do not affect the main features of the diagrams. Thus we have obtained a new approximate symmetry scheme which models the role of quadrupole interactions throughout a major oscillator shell that resembles the actual shell. As such we can imagine that it can be used to predict the evolution of observables that depend robustly on the number of nucleons interacting in a quadrupole field. Such predictions will obviously ignore other interactions, such as pairing, the roles of other shells, and the like, and it remains to be seen how that affects them. This is discussed again in a companion paper \cite{second}. Importantly, we remark that our approach is intended as a complement, not a replacement for, more comprehensive (and often computer-intensive) approaches such as large scale shell model calculations or multi-shell symplectic models.

\section{Conclusions} 

In this manuscript we propose that a proxy-SU(3) symmetry appears in heavy deformed nuclei, very similar to the Elliott SU(3) 
symmetry appearing in light (sd shell) nuclei. In order to demonstrate this fact, we use an elementary and completely transparent Nilsson calculation,
in which it becomes clear that the changes induced by replacing in each major shell the intruder parity Nilsson orbitals by their 0[110] counterparts 
are small, therefore offering the basis for a reliable approximate scheme. The main reasons behind the good quality of this approximation are:

1) the fact that the intruder parity orbitals have exactly the same orbital angular momentum, spin, and total angular momentum projection quantum numbers 
as their 0[110] substitutes, 

2) the small number and small contribution to the total Hamiltonian of the additional non-vanishing spin-orbit and angular-momentum-squared matrix elements appearing because 
of the approximation induced, which imply that the additional avoided crossings caused by the approximation are of small size, thus not affecting drastically the form of the Nilsson diagrams, 

3) because of 1) and 2), the real Nilsson diagrams have nearly the same structure as they would have had if the missing normal parity orbitals were present in the place of the intruder parity orbitals, completing an oscillator major shell with the appropriate U(N) symmetry algebra, having an SU(3) subalgebra.  

The main open question is if this proxy-SU(3) scheme can be of any practical use, in other words if the approximations made result in an SU(3) scheme from which reliable conclusions on physical quantities can be drawn. The demonstration that the Nilsson diagram based on the present proxy-SU(3) scheme is a good approximation to the actual one can be taken as a validation to use this scheme to carry out actual predictions for nuclear behavior. 

More specifically, the new set of states comprising the proxy
scheme now allows the system to be described by a symmetry
(instead of a collection of orbits that have to be solved in a
complex diagonalization process) corresponding to full sets of
oscillator states such as  the sdg orbits or the pfh orbits. 
Having a symmetry means that many results can now be obtained
analytically, often by inspection, and often in a parameter-free way.
This could involve, for example,  how various observables behave across a set of nuclei. 
It can also provide initial predictions for currently inaccessible nuclei. Ultimately, deviations from those predictions may help point to changes in shell structure or for the enhancement of certain interactions in unstable nuclei.
A first application is given in Ref. \cite{second}, in which it is shown that the present scheme can predict the prolate-over-oblate dominance in deformed nuclei, the location of the prolate-oblate shape phase transition in rare earth nuclei and specific predictions of the $\gamma$ and $\beta$ deformation values for deformed nuclei that are in good overall agreement with the data without any free parameters. We stress again that we do not view the proxy-SU(3) scheme as a substitute for detailed microscopic calculations or that it can even make plausible predictions for many of the spectroscopic results of such calculations.  But we do suggest that it can be a valuable, and certainly extremely simple, complement to such approaches, and a way of predicting certain more global properties of deformed nuclei related to their collectivity and shapes.

\section*{Appendix I} 

The spin-orbit term, ${\bf l} \cdot {\bf s}$, has diagonal matrix elements 
\begin{equation}
\langle n_z r s \Sigma |{\bf l} \cdot {\bf s} |  n_z r s \Sigma\rangle = (r-s) \Sigma = \Lambda \Sigma, 
\end{equation}
as well as non-diagonal matrix elements 
\begin{multline}
\langle n_z-1, r+1, s, \Sigma-1 |{\bf l} \cdot {\bf s} |  n_z r s \Sigma\rangle = -{1\over \sqrt{2}} \sqrt{n_z (r+1)}, \\
\langle n_z+1, r, s-1, \Sigma-1 |{\bf l} \cdot {\bf s} |  n_z r s \Sigma\rangle = {1\over \sqrt{2}} \sqrt{(n_z+1) s}, \\
\langle n_z+1, r-1, s, \Sigma+1 |{\bf l} \cdot {\bf s} |  n_z r s \Sigma\rangle = -{1\over \sqrt{2}} \sqrt{(n_z+1) r}, \\
\langle n_z-1, r, s+1, \Sigma+1 |{\bf l} \cdot {\bf s} |  n_z r s \Sigma\rangle = {1\over \sqrt{2}} \sqrt{n_z (s+1}. 
\end{multline}

The orbital angular momentum term, ${\bf l}^2$, has diagonal matrix elements 
\begin{equation}\label{l2diag}
\langle n_z r s \Sigma |{\bf l}^2 |  n_z r s \Sigma\rangle = 2 n_z(r+s+1) +(r+s)+(r-s)^2, 
\end{equation}
as well as non-diagonal matrix elements
\begin{multline} 
\langle n_z+2, r-1, s-1, \Sigma |{\bf l}^2 |  n_z r s \Sigma\rangle = -2\sqrt{(n_z+2)(n_z+1) r s }, \\
\langle n_z r s \Sigma |{\bf l}^2 |  n_z r s \Sigma\rangle =-2\sqrt{(n_z-1) n_z (r+1)(s+1)}.
\end{multline} 

The above equations are equivalent to these given in Ref. \cite{Nilsson2} in a $| n_z n_\perp \Lambda \Sigma \rangle$ notation. 
Additional results for matrix elements in various notations can be found in Refs. \cite{Rassey,Quentin,Boisson}. 

\section*{Supplementary material}

Tabulations of ${\bf l} \cdot {\bf s}$, ${\bf l}^2$, and $H$  matrix elements  for the 28--50 and pf, 82--126 and pfh, as well as 126--184 and sdgi shells as well as figures of spectra for the same shells, either involving only the diagonal terms of the Hamiltonian, or 
derived through the diagonalization of the full Hamiltonian, are provided in this section.

\begin{widetext}

\begin{table*}

\caption{${\bf l} \cdot {\bf s}$ matrix elements (in units of $\hbar \omega_0$) for Nilsson orbitals in the 28--50 shell (upper part) and 
 in the pf shell (lower part).}

\bigskip

\begin{tabular}{ r r r r r r r | r r r r r   }
                  &
${1\over 2}$[301] & ${1\over 2}$[321] & ${3\over 2}$[312] & ${1\over 2}$[310] & ${3\over 2}$[301] & ${5\over 2}$[303] & ${1\over 2}$[440] & 
${3\over 2}$[431] & ${5\over 2}$[422] & ${7\over 2}$[413] & ${9\over 2}[404]$ \\

\hline
    
 1/2[301] & $-0.5$ & 0        & 0        & $-1$ &  0     &              & 0        & 0     & 0     & 0     & 0     \\
 1/2[321] &         & $-0.5$ & 0        & 1    &  0     & 0            & 0        & 0     & 0     & 0     & 0     \\
 3/2[312] &         &         & $-1$ & 0        & 0.707 & 0            & 0        &  0    &  0    & 0     & 0     \\
 1/2[310] &  &     &         & 0        & 0      & 0            & 0        & 0     & 0     & 0     & 0     \\
 3/2[301] &         &        &    &       & 0.5  & 0            & 0        & 0     & 0     & 0     & 0     \\
 5/2[303] &         &         &         &         &       & $-1.5$     & 0        & 0     & 0     & 0     & 0     \\
 
 \hline 
 
 1/2[440] &         &        &         &         &       &             & 0        & 0     & 0     & 0     & 0     \\
 3/2[431] &        &         &         &         &       &             &         & 0.5 & 0     & 0     & 0     \\
 5/2[422] &         &        &         &         &       &             &         &      & 1 & 0     & 0     \\
 7/2[413] &         &         &         &         &       &             &         &      &      & 1.5 & 0     \\
 9/2[404] &       &        &         &         &       &             &         &      &      &    & 2 \\
 
 \hline
\end{tabular}

\vskip 0.3cm 

\begin{tabular}{ r r r r r r r  | r r r r   }
                  &
${1\over 2}$[301] & ${1\over 2}$[321] & ${3\over 2}$[312] & ${1\over 2}$[310] & ${3\over 2}$[301] & ${5\over 2}$[303] & ${1\over 2}$[330] & 
${3\over 2}$[321] & ${5\over 2}$[312] & ${7\over 2}$[303] \\ 

\hline
    
 1/2[301] & $-0.5$ & 0        & 0        & $-1$ &  0     &              & 0        & 0       & 0        & 0     \\
 1/2[321] &         & $-0.5$ & 0        & 1    &  0     & 0            & ${\bf -1.225}$& 0       & 0        & 0     \\
 3/2[312] &         &         & $-1$ & 0        & 0.707 & 0            & 0        &${\bf -1.414}$&  0       & 0     \\
 1/2[310] &  &    &         & 0        & 0      & 0            & 0        & 0       & 0        & 0     \\
 3/2[301] &      &         &    &         & 0.5  & 0            & 0        & 0       & 0        & 0     \\
 5/2[303] &        &        &         &         &      & $-1.5$     & 0        & 0       &${\bf -1.225}$ & 0     \\
 
 \hline 
 
 1/2[330] &        & &         &         &       &            & 0        & 0       & 0        & 0     \\
 3/2[321] &         &         &   &         &       &             &         & 0.5   & 0        & 0     \\
 5/2[312] &        &        &        &        &     &    &         &        & 1    & 0     \\
 7/2[303] &        &        &       &      &     &           &       &      &       & 1.5 \\

 \hline
\end{tabular}

\end{table*}

\begin{table*}

\caption{$H$ matrix elements (in units of $\hbar \omega_0$) with $\epsilon=0.3$ for Nilsson orbitals in the 28--50 shell (upper part) and
 in the pf shell (lower part).}

\bigskip

\begin{tabular}{ r r r r r r r | r r r r r   }
                  &
${1\over 2}$[301] & ${1\over 2}$[321] & ${3\over 2}$[312] & ${1\over 2}$[310] & ${3\over 2}$[301] & ${5\over 2}$[303] & ${1\over 2}$[440] & 
${3\over 2}$[431] & ${5\over 2}$[422] & ${7\over 2}$[413] & ${9\over 2}[404]$ \\

\hline
    
 1/2[301] & 4.88 & 0        & 0        & 0.16 &  0     &              & 0        & 0     & 0     & 0     & 0     \\
 1/2[321] &       & 4.28 & 0        & $-0.16$ &  0     & 0            & 0        & 0     & 0     & 0     & 0     \\
 3/2[312] &        &        & 4.66 & 0        & $-0.11$ & 0            & 0        &  0    &  0    & 0     & 0     \\
 1/2[310] & &     &        & 4.50     & 0      & 0            & 0        & 0     & 0     & 0     & 0     \\
 3/2[301] &        &       &  &         & 4.72  & 0            & 0        & 0     & 0     & 0     & 0     \\
 5/2[303] &        &         &         &        &     & 5.04     & 0        & 0     & 0     & 0     & 0     \\

 \hline 
 
 1/2[440] &      &       &       &        &      &            & 4.70        & 0     & 0     & 0     & 0     \\
 3/2[431] &         &         &         &         &       &             &         & 4.92 & 0     & 0     & 0     \\
 5/2[422] &         &         &         &         &       &             &         &      & 5.14 & 0     & 0     \\
 7/2[413] &         &         &         &         &       &             &         &      &      & 5.36 & 0     \\
 9/2[404] &         &        &         &         &       &             &         &      &      &      & 5.58 \\
 
 \hline
\end{tabular}

\vskip 0.3cm 

\begin{tabular}{ r r r r r r r | r r r r    }
                  &
${1\over 2}$[301] & ${1\over 2}$[321] & ${3\over 2}$[312] & ${1\over 2}$[310] & ${3\over 2}$[301] & ${5\over 2}$[303] & ${1\over 2}$[330] & 
${3\over 2}$[321] & ${5\over 2}$[312] & ${7\over 2}$[303] \\

\hline
    
 1/2[301] & 4.88 & 0        & 0        & 0.16 &  0     &              & 0        & 0     & 0     & 0         \\
 1/2[321] &         & 4.28 & 0        & $-0.16$ &  0     & 0            & {\bf 0.20}        & 0     & 0     & 0        \\
 3/2[312] &         &        & 4.66 & 0        & $-0.11$ & 0            & 0        &  {\bf 0.23}    &  0    & 0        \\
 1/2[310] & &     &        & 4.50     & 0      & 0            & 0        & 0     & 0     & 0        \\
 3/2[301] &         &         &  &         & 4.72  & 0            & 0        & 0     & 0     & 0        \\
 5/2[303] &        &        &        &        &     & 5.04     & 0        & 0     & {\bf 0.20}     & 0       \\

 \hline 
 
 1/2[330] &         &         &         &         &       &             & 4.70        & 0     & 0     & 0     \\
 3/2[321] &         &        &         &         &       &             &         & 4.92 & 0     & 0       \\
 5/2[312] &         &         &         &         &       &             &         &      & 5.14 & 0        \\
 7/2[303] &         &         &         &         &       &             &        &      &     & 5.36  \\
 
 \hline
\end{tabular}

\end{table*}

\end{widetext}

\begin{turnpage}

\begingroup

\squeezetable 

\begin{table*}

\caption{${\bf l} \cdot {\bf s}$ matrix elements (in units of $\hbar \omega_0$) for Nilsson orbitals in the 82--126 shell (upper part) and 
 in the full pfh shell (lower part).}

\bigskip

\begin{tabular}{ r r r r r r r r r r r r r r r r | r r r r r r r }
                  &
$\frac{1}{2}$[501] & $\frac{1}{2}$[521] & $\frac{3}{2}$[512] & $\frac{1}{2}$[510] & $\frac{3}{2}$[501] & $\frac{5}{2}$[503] & $\frac{1}{2}$[541] & 
$\frac{3}{2}$[532] & $\frac{5}{2}$[523] & $\frac{7}{2}$[514] & $\frac{1}{2}$[530] & $\frac{3}{2}$[521] & $\frac{5}{2}$[512] & $\frac{7}{2}$[503] & 
$\frac{9}{2}$[505] 
& $\frac{1}{2}$[660] & $\frac{3}{2}$[651] & $\frac{5}{2}$[642] & $\frac{7}{2}$[633] & $\frac{9}{2}$[624] & 
$\frac{11}{2}$[615] & $\frac{13}{2}$[606] 
\\

\hline

1/2[501] & $-0.5$ & 0 & 0 & $-1.225$ & 0  & 0  & 0  & 0  & 0  & 0  & 0  & 0  & 0  & 0 & 0 & 0  & 0 & 0 & 0 & 0 &  0 & 0 \\
1/2[521] &  & $-0.5$ & 0 & 1.414 & 0  & 0  & 0  & 0  & 0  & 0  &$-1.732$  & 0  & 0  & 0 & 0 & 0  & 0 & 0 & 0 & 0 & 0 & 0 \\
3/2[512] &  &  & $-1$ & 0 & 1  & 0  & 0  & 0  & 0  & 0  & 0  & $-1.732$ & 0  & 0 & 0 & 0  & 0 & 0 & 0 & 0 & 0 & 0 \\
1/2[510] &  &  &  & 0 & 0  & 0  & 0  & 0  & 0  & 0  & 0  & 0  & 0  & 0 & 0 & 0  & 0 &  0 & 0 & 0 & 0 & 0 \\
3/2[501] &     &     &     &     & 0.5 & 0 & 0 & 0 & 0 & 0 & 0 & 0 & 0 & 0    & 0    & 0 & 0 & 0   & 0 & 0 & 0   & 0 \\
5/2[503] &     &     &     &    & & $-1.5$ & 0 & 0 & 0 & 0 & 0 & 0 & $-1.414$ & 0    & 0    & 0 & 0 & 0   & 0 & 0 & 0   & 0 \\
1/2[541] &   &    &    &    & &  & $-0.5$ & 0 & 0 & 0  & 1.414 & 0 & 0 & 0    & 0    & 0 & 0 & 0   & 0 & 0 & 0   & 0 \\
\hline
\end{tabular}

\vskip 0.3cm

\begin{tabular}{ r r r r r r r r r r r r r r r r | r r r r r r  }
       &  
$\frac{1}{2}$[501] & $\frac{1}{2}$[521] & $\frac{3}{2}$[512] & $\frac{1}{2}$[510] & $\frac{3}{2}$[501] & $\frac{5}{2}$[503] & $\frac{1}{2}$[541] & 
$\frac{3}{2}$[532] & $\frac{5}{2}$[523] & $\frac{7}{2}$[514] & $\frac{1}{2}$[530] & $\frac{3}{2}$[521] & $\frac{5}{2}$[512] & $\frac{7}{2}$[503] & 
$\frac{9}{2}$[505]   
& $\frac{1}{2}$[550] & $\frac{3}{2}$[541] & $\frac{5}{2}$[532] & $\frac{7}{2}$[523] & $\frac{9}{2}$[514] & 
$\frac{11}{2}$[505] 
\\

\hline

1/2[501] & $-0.5$ & 0 & 0 & $-1.225$ & 0  & 0  & 0  & 0  & 0  & 0  & 0  & 0  & 0  & 0 & 0 & 0  & 0 & 0 & 0 & 0 &  0  \\
1/2[521] &  & $-0.5$ & 0 & 1.414 & 0  & 0  & 0  & 0  & 0  & 0  &$-1.732$  & 0  & 0  & 0 & 0 & 0  & 0 & 0 & 0 & 0 & 0  \\
3/2[512] &  &  & $-1$ & 0 & 1  & 0  & 0  & 0  & 0  & 0  & 0  & $-1.732$ & 0  & 0 & 0 & 0  & 0 & 0 & 0 & 0 & 0  \\
1/2[510] &  &  &  & 0 & 0  & 0  & 0  & 0  & 0  & 0  & 0  & 0  & 0  & 0 & 0 & 0  & 0 &  0 & 0 & 0 & 0  \\
3/2[501] &     &     &    &   & 0.5 & 0 & 0 & 0 & 0 & 0 & 0 & 0 & 0 & 0    & 0    & 0 & 0 & 0   & 0 & 0 & 0   \\
5/2[503] &    &   &   &   &  & $-1.5$ & 0 & 0 & 0 & 0 & 0 & 0 & $-1.414$ & 0    & 0    & 0 & 0 & 0   & 0 & 0 & 0    \\
1/2[541] &   &    &     &     &  &  & $-0.5$ & 0 & 0 & 0  & 1.414 & 0 & 0 & 0    & 0    & ${\bf -1.581}$ & 0 & 0   & 0 & 0 & 0  \\
\hline
\end{tabular}

\end{table*}

\endgroup

\end{turnpage}

\begin{turnpage}

\begingroup

\squeezetable 

\begin{table*}

\caption{${\bf l}^2$ matrix elements (in units of $\hbar \omega_0$) for Nilsson orbitals in the 82--126 shell (upper part) and
 in the full pfh shell (lower part).}

\bigskip

\begin{tabular}{ r r r r r r r r r r r r r r r r | r r r r r r r }
                  &
${1\over 2}$[501] & ${1\over 2}$[521] & ${3\over 2}$[512] & ${1\over 2}$[510] & ${3\over 2}$[501] & ${5\over 2}$[503] & ${1\over 2}$[541] & 
${3\over 2}$[532] & ${5\over 2}$[523] & ${7\over 2}$[514] & ${1\over 2}$[530] & ${3\over 2}$[521] & ${5\over 2}$[512] & ${7\over 2}$[503] & 
${9\over 2}$[505] & ${1\over 2}$[660] & ${3\over 2}$[651] & ${5\over 2}$[642] & ${7\over 2}$[633] & ${9\over 2}$[624] & 
${11\over 2}$[615] & ${13\over 2}$[606] 
\\

\hline

1/2[501] & 6 & $-6.928$ & 0 & 0 & 0  & 0  & 0  & 0  & 0  & 0  & 0  & 0  & 0  & 0 & 0 & 0  & 0 & 0 & 0 & 0 &  0 & 0 \\
1/2[521] &  & 20 & 0 & 0 & 0  & 0  & $-9.798$  & 0  & 0  & 0  & 0  & 0  & 0  & 0 & 0 & 0  & 0 & 0 & 0 & 0 & 0 & 0 \\
3/2[512] &  &  & 18 & 0 & 0  & 0  & 0  & $-8.485$ & 0 & 0  & 0  & 0  & 0  & 0 & 0 & 0  & 0 & 0 & 0 & 0 & 0 & 0 \\
1/2[510] &  & &  & 14 & 0  & 0  & 0  & 0  & 0  & 0  & $-9.798$  & 0  & 0  & 0 & 0 & 0  & 0 &  0 & 0 & 0 & 0 & 0 \\
3/2[501] &     &    &    &    & 6 & 0 & 0 & 0 & 0 & 0 & 0 & -6.928 & 0 & 0    & 0    & 0 & 0 & 0   & 0 & 0 & 0   & 0 \\

5/2[503] &     &    &     &     &  & 14 & 0 & 0 & $-5.657$ & 0 & 0 & 0 & 0 & 0    & 0    & 0 & 0 & 0   & 0 & 0 & 0   & 0 \\
1/2[541] &     & &     &     &  &  & 18 & 0 & 0 & 0  & 0 & 0 & 0 & 0    & 0    & 0 & 0 & 0   & 0 & 0 & 0   & 0 \\
\hline
\end{tabular}

\vskip 0.3cm 

\begin{tabular}{ r r r r r r r r r r r r r r r r | r r r r r r  }
                  &
${1\over 2}$[501] & ${1\over 2}$[521] & ${3\over 2}$[512] & ${1\over 2}$[510] & ${3\over 2}$[501] & ${5\over 2}$[503] & ${1\over 2}$[541] & 
${3\over 2}$[532] & ${5\over 2}$[523] & ${7\over 2}$[514] & ${1\over 2}$[530] & ${3\over 2}$[521] & ${5\over 2}$[512] & ${7\over 2}$[503] & 
${9\over 2}$[505] & ${1\over 2}$[550] & ${3\over 2}$[541] & ${5\over 2}$[532] & ${7\over 2}$[523] & ${9\over 2}$[514] & ${11\over 2}$[505]
\\

\hline

1/2[501] & 6 & $-6.928$ & 0 & 0 & 0  & 0  & 0  & 0  & 0  & 0  & 0  & 0  & 0  & 0 & 0 & 0  & 0 & 0 & 0 & 0 &  0 \\
1/2[521] &  & 20 & 0 & 0 & 0  & 0  & $-9.798$  & 0  & 0  & 0  & 0  & 0  & 0  & 0 & 0 & 0  & 0 & 0 & 0 & 0 & 0  \\
3/2[512] &  &  & 18 & 0 & 0  & 0  & 0 & $-8.485$ & 0 & 0  & 0  & 0  & 0  & 0 & 0 & 0  & 0 & 0 & 0 & 0 & 0 \\
1/2[510] &  &  &  & 14 & 0  & 0  & 0  & 0  & 0  & 0  & $-9.798$  & 0  & 0  & 0 & 0 & 0  & 0 &  0 & 0 & 0 & 0  \\
3/2[501] &     &     &     &     & 6 & 0 & 0 & 0 & 0 & 0 & 0 & -6.928 & 0 & 0    & 0    & 0 & 0 & 0   & 0 & 0 & 0  \\

5/2[503] &    &     &    &     &  & 14 & 0 & 0 & $-5.657$ & 0 & 0 & 0 & 0 & 0    & 0    & 0 & 0 & 0   & 0 & 0 & 0  \\
1/2[541] &   &  &     &     &  &  & 18 & 0 & 0 & 0  & 0 & 0 & 0 & 0    & 0    & 0 & 0 & 0   & 0 & 0 & 0  \\
\hline
\end{tabular}

\end{table*}

\endgroup

\end{turnpage}

\begin{turnpage}

\begingroup

\squeezetable 

\begin{table*}

\caption{$H$ matrix elements (in units of $\hbar \omega_0$) with $\epsilon=0.3$ for Nilsson orbitals in the 82--126 neutron shell (upper part)
and in the full pfh neutron shell (lower part).}

\bigskip

\begin{tabular}{ r r r r r r r r r r r r r r r r | r r r r r r r }
                  &
${1\over 2}$[501] & ${1\over 2}$[521] & ${3\over 2}$[512] & ${1\over 2}$[510] & ${3\over 2}$[501] & ${5\over 2}$[503] & ${1\over 2}$[541] & 
${3\over 2}$[532] & ${5\over 2}$[523] & ${7\over 2}$[514] & ${1\over 2}$[530] & ${3\over 2}$[521] & ${5\over 2}$[512] & ${7\over 2}$[503] & 
${9\over 2}$[505] & ${1\over 2}$[660] & ${3\over 2}$[651] & ${5\over 2}$[642] & ${7\over 2}$[633] & ${9\over 2}$[624] & 
${11\over 2}$[615] & ${13\over 2}$[606] 
\\

\hline

1/2[501] & 7.44 & 0.19 & 0 & 0.16 & 0  & 0  & 0  & 0  & 0  & 0  & 0  & 0  & 0  & 0 & 0 & 0  & 0 & 0 & 0 & 0 &  0 & 0 \\
1/2[521] &  & 6.46 & 0 & $-0.18$ & 0  & 0  & 0.26  & 0  & 0  & 0  & 0.22  & 0  & 0  & 0 & 0 & 0  & 0 & 0 & 0 & 0 & 0 & 0 \\
3/2[512] &  &  & 6.88 & 0 & $-0.13$  & 0  & 0  & 0.23  & 0  & 0  & 0  & 0.22 & 0  & 0 & 0 & 0  & 0 & 0 & 0 & 0 & 0 & 0 \\
1/2[510] & &  &  & 6.86 & 0  & 0  & 0  & 0  & 0  & 0  & 0.26  & 0  & 0  & 0 & 0 & 0  & 0 &  0 & 0 & 0 & 0 & 0 \\
3/2[501] &    &    &    &    & 7.31 & 0 & 0 & 0 & 0 & 0 & 0 & 0.19 & 0 & 0    & 0    & 0 & 0 & 0   & 0 & 0 & 0   & 0 \\
5/2[503] &     &     &     &     &  & 7.35 & 0 & 0 & 0.15 & 0 & 0 & 0 & 0.18 & 0    & 0    & 0 & 0 & 0   & 0 & 0 & 0   & 0 \\
1/2[541] &   &  &   &   &  & & 5.92 & 0 & 0 & 0  & $-0.18$ & 0 & 0 & 0    & 0    & 0 & 0 & 0   & 0 & 0 & 0   & 0 \\
\hline
\end{tabular}

\vskip 0.3cm 

\begin{tabular}{ r r r r r r r r r r r r r r r r | r r r r r r }
                  &
${1\over 2}$[501] & ${1\over 2}$[521] & ${3\over 2}$[512] & ${1\over 2}$[510] & ${3\over 2}$[501] & ${5\over 2}$[503] & ${1\over 2}$[541] & 
${3\over 2}$[532] & ${5\over 2}$[523] & ${7\over 2}$[514] & ${1\over 2}$[530] & ${3\over 2}$[521] & ${5\over 2}$[512] & ${7\over 2}$[503] & 
${9\over 2}$[505] & ${1\over 2}$[550] & ${3\over 2}$[541] & ${5\over 2}$[532] & ${7\over 2}$[523] & ${9\over 2}$[514] & 
${11\over 2}$[505] 
\\

\hline

1/2[501] & 7.44 & 0.19 & 0 & 0.16 & 0  & 0  & 0  & 0  & 0  & 0  & 0  & 0  & 0  & 0 & 0 & 0  & 0 & 0 & 0 & 0 &  0  \\
1/2[521] &  & 6.46 & 0 & $-0.18$ & 0  & 0  & 0.26  & 0  & 0  & 0  & 0.22  & 0  & 0  & 0 & 0 & 0  & 0 & 0 & 0 & 0 & 0  \\
3/2[512] &  &  & 6.88 & 0 & $-0.13$  & 0  & 0  & 0.23  & 0  & 0  & 0  & 0.22 & 0  & 0 & 0 & 0  & 0 & 0 & 0 & 0 & 0  \\
1/2[510] & & &  & 6.86 & 0  & 0  & 0  & 0  & 0  & 0  & 0.26  & 0  & 0  & 0 & 0 & 0  & 0 &  0 & 0 & 0 & 0 \\
3/2[501] &  & &  &  & 7.31 & 0 & 0 & 0 & 0 & 0 & 0 & 0.19 & 0 & 0    & 0    & 0 & 0 & 0   & 0 & 0 & 0  \\
5/2[503] &  &  &  &  &  & 7.35 & 0 & 0 & 0.15 & 0 & 0 & 0 & 0.18 & 0    & 0    & 0 & 0 & 0   & 0 & 0 & 0  \\
1/2[541] &  &   &   &   &  &  & 5.92 & 0 & 0 & 0  & $-0.18$ & 0 & 0 & 0    & 0    & {\bf 0.20} & 0 & 0   & 0 & 0 & 0  \\
\hline
\end{tabular}

\end{table*}

\endgroup

\end{turnpage}

\begin{turnpage}

\begingroup

\squeezetable 

\begin{table*}

\caption{$H$ matrix elements (in units of $\hbar \omega_0$) with  $\epsilon=0.3$ for Nilsson orbitals in the 82--126 proton shell (upper part)
and in the full pfh proton shell (lower part).}

\bigskip

\begin{tabular}{ r r r r r r r r r r r r r r r r | r r r r r r r }
                  &
${1\over 2}$[501] & ${1\over 2}$[521] & ${3\over 2}$[512] & ${1\over 2}$[510] & ${3\over 2}$[501] & ${5\over 2}$[503] & ${1\over 2}$[541] & 
${3\over 2}$[532] & ${5\over 2}$[523] & ${7\over 2}$[514] & ${1\over 2}$[530] & ${3\over 2}$[521] & ${5\over 2}$[512] & ${7\over 2}$[503] & 
${9\over 2}$[505] & ${1\over 2}$[660] & ${3\over 2}$[651] & ${5\over 2}$[642] & ${7\over 2}$[633] & ${9\over 2}$[624] & 
${11\over 2}$[615] & ${13\over 2}$[606] 
\\

\hline

1/2[501] & 7.58 & 0.26 & 0 & 0.14 & 0  & 0  & 0  & 0  & 0  & 0  & 0  & 0  & 0  & 0 & 0 & 0  & 0 & 0 & 0 & 0 &  0 & 0 \\
1/2[521] &  & 6.46 & 0 & $-0.16$ & 0  & 0  & 0.37  & 0  & 0  & 0  & 0.20  & 0  & 0  & 0 & 0 & 0  & 0 & 0 & 0 & 0 & 0 & 0 \\
3/2[512] &  &  & 6.89 & 0 & $-0.12$  & 0  & 0  & 0.32  & 0  & 0  & 0  & 0.20 & 0  & 0 & 0 & 0  & 0 & 0 & 0 & 0 & 0 & 0 \\
1/2[510] &  &  &  & 6.93 & 0  & 0  & 0  & 0  & 0  & 0  & 0.37  & 0  & 0  & 0 & 0 & 0  & 0 &  0 & 0 & 0 & 0 & 0 \\
3/2[501] & & &  & & 7.47 & 0 & 0 & 0 & 0 & 0 & 0 & 0.26 & 0 & 0    & 0    & 0 & 0 & 0   & 0 & 0 & 0   & 0 \\
5/2[503] & & & & &  & 7.40 & 0 & 0 & 0.21 & 0 & 0 & 0 & 0.16 & 0    & 0    & 0 & 0 & 0   & 0 & 0 & 0   & 0 \\
1/2[541] & &  &  &  &  &  & 5.93 & 0 & 0 & 0  & $-0.16$ & 0 & 0 & 0    & 0    & 0 & 0 & 0   & 0 & 0 & 0   & 0 \\
\hline
\end{tabular}

\vskip 0.3cm 

\begin{tabular}{ r r r r r r r r r r r r r r r r | r r r r r r }
                  &
${1\over 2}$[501] & ${1\over 2}$[521] & ${3\over 2}$[512] & ${1\over 2}$[510] & ${3\over 2}$[501] & ${5\over 2}$[503] & ${1\over 2}$[541] & 
${3\over 2}$[532] & ${5\over 2}$[523] & ${7\over 2}$[514] & ${1\over 2}$[530] & ${3\over 2}$[521] & ${5\over 2}$[512] & ${7\over 2}$[503] & 
${9\over 2}$[505] & ${1\over 2}$[550] & ${3\over 2}$[541] & ${5\over 2}$[532] & ${7\over 2}$[523] & ${9\over 2}$[514] & 
${11\over 2}$[505] 
\\

\hline

1/2[501] & 7.58 & 0.26 & 0 & 0.14 & 0  & 0  & 0  & 0  & 0  & 0  & 0  & 0  & 0  & 0 & 0 & 0  & 0 & 0 & 0 & 0 &  0  \\
1/2[521] &  & 6.46 & 0 & $-0.16$ & 0  & 0  & 0.37  & 0  & 0  & 0  & 0.20  & 0  & 0  & 0 & 0 & 0  & 0 & 0 & 0 & 0 & 0  \\
3/2[512] &  &  & 6.89 & 0 & $-0.12$  & 0  & 0  & 0.32  & 0  & 0  & 0  & 0.20 & 0  & 0 & 0 & 0  & 0 & 0 & 0 & 0 & 0  \\
1/2[510] &  &  &  & 6.93 & 0  & 0  & 0  & 0  & 0  & 0  & 0.37  & 0  & 0  & 0 & 0 & 0  & 0 &  0 & 0 & 0 & 0 \\
3/2[501] & & & & & 7.47 & 0 & 0 & 0 & 0 & 0 & 0 & 0.26 & 0 & 0    & 0    & 0 & 0 & 0   & 0 & 0 & 0  \\
5/2[503] & & & & &  & 7.40 & 0 & 0 & 0.21 & 0 & 0 & 0 & 0.16 & 0    & 0    & 0 & 0 & 0   & 0 & 0 & 0  \\
1/2[541] & &  & &  &  &  & 5.93 & 0 & 0 & 0  & $-0.16$ & 0 & 0 & 0    & 0    & {\bf 0.18} & 0 & 0   & 0 & 0 & 0  \\
\hline
\end{tabular}

\end{table*}

\endgroup

\end{turnpage}

\begin{turnpage}


\begingroup

\squeezetable

\begin{table*}

\caption{${\bf l} \cdot {\bf s}$ matrix elements (in units of $\hbar \omega_0$) for Nilsson orbitals in the 126--184 shell (upper part) and 
in the full sdgi shell (lower part).}

\bigskip
\tabcolsep=1pt
\begin{tabular}{ r r r r r r r r r r r r r r r r r r r r r r | r r r r r r r r }
                  &
${1\over 2}$[611] & ${1\over 2}$[600] & ${3\over 2}$[602] & ${1\over 2}$[631] & ${3\over 2}$[622] & ${5\over 2}$[613] & ${1\over 2}$[620] & 
${3\over 2}$[611] & ${5\over 2}$[602] & ${7\over 2}$[604] & ${1\over 2}$[651] & ${3\over 2}$[642] & ${5\over 2}$[633] & ${7\over 2}$[624] & 
${9\over 2}$[615] & ${1\over 2}$[640] & ${3\over 2}$[631] & ${5\over 2}$[622] & ${7\over 2}$[613] & ${9\over 2}$[604] & ${11\over 2}$[606]& 
${1\over 2}$[770] & ${3\over 2}$[761] & ${5\over 2}$[752] & ${7\over 2}$[743] & ${9\over 2}$[734] &${11\over 2}$[725] & ${13\over 2}$[716]& 
${15\over 2}$[707] \\

\hline

1/2[611]& $-0.5$ & 1.225 & 0 & 0 & 0  & 0  & $-1.732$ & 0  & 0  & 0  & 0  & 0  & 0  & 0 & 0 & 0  & 0 & 0 & 0 & 0 &  0 & 0 & 0 & 0 & 0 & 0 & 0 & 0 & 0 \\
1/2[600]& & 0 & 0 & 0 & 0  & 0  & 0  & 0  & 0  & 0  & 0  & 0  & 0  & 0 & 0 & 0  & 0 & 0 & 0 & 0 & 0 & 0& 0 & 0 & 0 & 0 & 0 & 0 & 0  \\
3/2[602] & & & $-1$ & 0 & 0  & 0  & 0  & $-1.414$  & 0  & 0  & 0  & 0 & 0  & 0 & 0 & 0  & 0 & 0 & 0 & 0 & 0 & 0& 0 & 0 & 0 & 0 & 0 & 0 & 0  \\
1/2[631] &  &  &  & $-0.5$ & 0 & 0 & 1.732 & 0 & 0 & 0  & 0  & 0  & 0  & 0 & 0 & $-2$ & 0 &  0 & 0 & 0 & 0 & 0& 0 & 0 & 0 & 0 & 0 & 0 & 0  \\
3/2[622] &  &  &  &  & $-1$ & 0 & 0 & 1.414 & 0 & 0 & 0 & 0 & 0 & 0 & 0 & 0 & $-2.121$ & 0 & 0 & 0 & 0 & 0& 0 & 0 & 0 & 0 & 0 & 0 & 0  \\

5/2[613] &  &  &  &  &  & $-1.5$ & 0 & 0 & 1 & 0 & 0 & 0 & 0 & 0 & 0 & 0 & 0 & $-2$ & 0 & 0 & 0 & 0& 0 & 0 & 0 & 0 & 0 & 0 & 0  \\
1/2[620] &  &  &  &  &  &  & 0  & 0 & 0 & 0  & 0 & 0 & 0 & 0 & 0 & 0 & 0 & 0 & 0 & 0 & 0& 0& 0 & 0 & 0 & 0 & 0 & 0 & 0  \\
\hline
\end{tabular}

\vskip 0.3cm 

\begin{tabular}{ r r r r r r r r r r r r r r r r r r r r r r | r r r r r r r  }
                  &
${1\over 2}$[611] & ${1\over 2}$[600] & ${3\over 2}$[602] & ${1\over 2}$[631] & ${3\over 2}$[622] & ${5\over 2}$[613] & ${1\over 2}$[620] & 
${3\over 2}$[611] & ${5\over 2}$[602] & ${7\over 2}$[604] & ${1\over 2}$[651] & ${3\over 2}$[642] & ${5\over 2}$[633] & ${7\over 2}$[624] & 
${9\over 2}$[615] & ${1\over 2}$[640] & ${3\over 2}$[631] & ${5\over 2}$[622] & ${7\over 2}$[613] & ${9\over 2}$[604] & ${11\over 2}$[606]& 
${1\over 2}$[660] & ${3\over 2}$[651] & ${5\over 2}$[642] & ${7\over 2}$[633] & ${9\over 2}$[624] &${11\over 2}$[615] & ${13\over 2}$[606]\\

\hline

1/2[611]& $-0.5$ & 1.225 & 0 & 0 & 0  & 0  & $-1.732$ & 0  & 0  & 0  & 0  & 0  & 0  & 0 & 0 & 0  & 0 & 0 & 0 & 0 &  0 & 0 & 0 & 0 & 0 & 0 & 0 & 0 \\
1/2[600]& & 0 & 0 & 0 & 0  & 0  & 0  & 0  & 0  & 0  & 0  & 0  & 0  & 0 & 0 & 0  & 0 & 0 & 0 & 0 & 0 & 0& 0 & 0 & 0 & 0 & 0 & 0  \\
3/2[602]& & & $-1$ & 0 & 0 & 0 & 0& $-1.414$& 0  & 0  & 0  & 0 & 0  & 0 & 0 & 0  & 0 & 0 & 0 & 0 & 0 & 0& 0 & 0 & 0 & 0 & 0 & 0 \\
1/2[631] & & & & $-0.5$ & 0& 0& 1.732& 0& 0 & 0  & 0  & 0  & 0  & 0 & 0 & $-2$ & 0 &  0 & 0 & 0 & 0 & 0& 0 & 0 & 0 & 0 & 0 & 0  \\
3/2[622] & & & & & $-1$& 0& 0& 1.414 & 0 & 0 & 0 & 0 & 0 & 0 & 0 & 0 & $-2.121$ & 0   & 0 & 0 & 0   & 0& 0 & 0 & 0 & 0 & 0 & 0  \\

5/2[613] & & & & & & $-1.5$ & 0 & 0 & 1 & 0 & 0 & 0 & 0 & 0 & 0 & 0 & 0 & $-2$ & 0 & 0 & 0 & 0& 0 & 0 & 0 & 0 & 0 & 0  \\
1/2[620] &  &  &  &  & &  & 0  & 0 & 0 & 0  & 0 & 0 & 0 & 0 & 0 & 0 & 0 & 0 & 0 & 0 & 0& 0& 0 & 0 & 0 & 0 & 0 & 0  \\
\hline
\end{tabular}

\end{table*}

\endgroup

\end{turnpage}

\begin{turnpage}


\begingroup

\squeezetable

\begin{table*}

\caption{${\bf l}^2$ matrix elements (in units of $\hbar \omega_0$) for Nilsson orbitals in the 126--184 shell (upper part) and 
in the full sdgi shell (lower part).}

\bigskip
\tabcolsep=1pt
\begin{tabular}{ r r r r r r r r r r r r r r r r r r r r r r | r r r r r r r r }
                  &
${1\over 2}$[611] & ${1\over 2}$[600] & ${3\over 2}$[602] & ${1\over 2}$[631] & ${3\over 2}$[622] & ${5\over 2}$[613] & ${1\over 2}$[620] & 
${3\over 2}$[611] & ${5\over 2}$[602] & ${7\over 2}$[604] & ${1\over 2}$[651] & ${3\over 2}$[642] & ${5\over 2}$[633] & ${7\over 2}$[624] & 
${9\over 2}$[615] & ${1\over 2}$[640] & ${3\over 2}$[631] & ${5\over 2}$[622] & ${7\over 2}$[613] & ${9\over 2}$[604] & ${11\over 2}$[606]& 
${1\over 2}$[770] & ${3\over 2}$[761] & ${5\over 2}$[752] & ${7\over 2}$[743] & ${9\over 2}$[734] &${11\over 2}$[725] & ${13\over 2}$[716]& 
${15\over 2}$[707] \\

\hline

1/2[611]& 18 & 0 & 0 & $-12$ & 0  & 0  &0 & 0  & 0  & 0  & 0  & 0  & 0  & 0 & 0 & 0  & 0 & 0 & 0 & 0 &  0 & 0 & 0 & 0 & 0 & 0 & 0 & 0 & 0 \\
1/2[600]&  & 6 & 0 & 0 & 0  & 0  & $-8.485$  & 0  & 0  & 0  & 0  & 0  & 0  & 0 & 0 & 0  & 0 & 0 & 0 & 0 & 0 & 0& 0 & 0 & 0 & 0 & 0 & 0 & 0  \\
3/2[602] &  &  & 10 & 0 & 0  & 0  & 0  & 0  & 0  & 0  & 0  & 0 & 0  & 0 & 0 & 0  & 0 & 0 & 0 & 0 & 0 & 0& 0 & 0 & 0 & 0 & 0 & 0 & 0  \\
1/2[631] &  &  &  & 28 & 0  & 0  & 0  & 0  & 0  & 0  & $-12.649$  & 0  & 0  & 0 & 0 & 0 & 0 &  0 & 0 & 0 & 0 & 0& 0 & 0 & 0 & 0 & 0 & 0 & 0  \\
3/2[622] &  &  &  &  & 28 & 0 & 0 & 0 & 0 & 0 & 0 & 0 & 0 & 0 & 0 & 0 & 0 & 0   & 0 & 0 & 0   & 0& 0 & 0 & 0 & 0 & 0 & 0 & 0  \\

5/2[613] & & & & & & 26 & 0 & 0 & 0 & 0 & 0 & 0 & $-9.798$ & 0 & 0 & 0 & 0 & 0 & 0 & 0 & 0 & 0& 0 & 0 & 0 & 0 & 0 & 0 & 0  \\
1/2[620] & & & & & & & 24  & 0 & 0 & 0  & 0 & 0 & 0 & 0 & 0 & $-13.856$ & 0 & 0 & 0 & 0 & 0& 0& 0 & 0 & 0 & 0 & 0 & 0 & 0  \\
\hline
\end{tabular}

\vskip 0.3cm

\begin{tabular}{ r r r r r r r r r r r r r r r r r r r r r r | r r r r r r r  }
                  &
${1\over 2}$[611] & ${1\over 2}$[600] & ${3\over 2}$[602] & ${1\over 2}$[631] & ${3\over 2}$[622] & ${5\over 2}$[613] & ${1\over 2}$[620] & 
${3\over 2}$[611] & ${5\over 2}$[602] & ${7\over 2}$[604] & ${1\over 2}$[651] & ${3\over 2}$[642] & ${5\over 2}$[633] & ${7\over 2}$[624] & 
${9\over 2}$[615] & ${1\over 2}$[640] & ${3\over 2}$[631] & ${5\over 2}$[622] & ${7\over 2}$[613] & ${9\over 2}$[604] & ${11\over 2}$[606]& 
${1\over 2}$[660] & ${3\over 2}$[651] & ${5\over 2}$[642] & ${7\over 2}$[633] & ${9\over 2}$[624] &${11\over 2}$[615] & ${13\over 2}$[606] \\

\hline

1/2[611]& 18 & 0 & 0& $-12$ & 0 & 0 &0 & 0& 0 & 0  & 0  & 0  & 0  & 0 & 0 & 0  & 0 & 0 & 0 & 0 &  0 & 0 & 0 & 0 & 0 & 0 & 0 & 0 \\
1/2[600]& & 6 & 0 & 0 & 0 & 0 & $-8.485$& 0 & 0 & 0 & 0  & 0  & 0  & 0 & 0 & 0  & 0 & 0 & 0 & 0 & 0 & 0& 0 & 0 & 0 & 0 & 0 & 0  \\
3/2[602]& & & 10 & 0 & 0  & 0  & 0  & 0  & 0  & 0  & 0  & 0 & 0  & 0 & 0 & 0  & 0 & 0 & 0 & 0 & 0 & 0& 0 & 0 & 0 & 0 & 0 & 0  \\
1/2[631]& & & & 28 & 0 & 0 & 0& 0& 0  & 0  & $-12.649$  & 0  & 0  & 0 & 0 & 0 & 0 &  0 & 0 & 0 & 0 & 0& 0 & 0 & 0 & 0 & 0 & 0   \\
3/2[622]& & & & & 28 & 0 & 0 & 0 & 0 & 0 & 0 & 0 & 0 & 0 & 0 & 0 & 0 & 0   & 0 & 0 & 0   & 0& 0 & 0 & 0 & 0 & 0 & 0   \\

5/2[613]& & & & & & 26 & 0 & 0 & 0 & 0 & 0 & 0 & $-9.798$ & 0 & 0 & 0 & 0 & 0 & 0 & 0 & 0 & 0& 0 & 0 & 0 & 0 & 0 & 0  \\
1/2[620]& & & & & & & 24  & 0 & 0 & 0  & 0 & 0 & 0 & 0 & 0 & $-13.856$ & 0 & 0 & 0 & 0 & 0& 0& 0 & 0 & 0 & 0 & 0 & 0   \\
\hline
\end{tabular}

\end{table*}

\endgroup

\end{turnpage}

\begin{turnpage}


\begingroup

\squeezetable

\begin{table*}

\caption{$H$ matrix elements (in units of $\hbar \omega_0$) with $\epsilon=0.3$ for Nilsson orbitals in the 126--184 shell (upper part) 
and in the full sdgi shell (lower part).}

\bigskip
\tabcolsep=1pt
\begin{tabular}{ r r r r r r r r r r r r r r r r r r r r r r | r r r r r r r r }
                  &
${1\over 2}$[611] & ${1\over 2}$[600] & ${3\over 2}$[602] & ${1\over 2}$[631] & ${3\over 2}$[622] & ${5\over 2}$[613] & ${1\over 2}$[620] & 
${3\over 2}$[611] & ${5\over 2}$[602] & ${7\over 2}$[604] & ${1\over 2}$[651] & ${3\over 2}$[642] & ${5\over 2}$[633] & ${7\over 2}$[624] & 
${9\over 2}$[615] & ${1\over 2}$[640] & ${3\over 2}$[631] & ${5\over 2}$[622] & ${7\over 2}$[613] & ${9\over 2}$[604] & ${11\over 2}$[606]& 
${1\over 2}$[770] & ${3\over 2}$[761] & ${5\over 2}$[752] & ${7\over 2}$[743] & ${9\over 2}$[734] &${11\over 2}$[725] & ${13\over 2}$[716]& 
${15\over 2}$[707] \\

\hline

1/2[611]& 8.05 & $-0.16$ & 0 & 0.25 & 0 & 0 & 0.22 & 0 & 0 & 0 & 0 & 0 & 0  & 0 & 0 & 0  & 0 & 0 & 0 & 0 &  0 & 0 & 0 & 0 & 0 & 0 & 0 & 0 & 0 \\
1/2[600]& & 8.53 & 0 & 0 & 0  & 0  & 0.17  & 0  & 0  & 0  & 0  & 0  & 0  & 0 & 0 & 0  & 0 & 0 & 0 & 0 & 0 & 0& 0 & 0 & 0 & 0 & 0 & 0 & 0  \\
3/2[602] & & & 8.58 & 0 & 0  & 0  & 0  & 0.18  & 0  & 0  & 0  & 0 & 0  & 0 & 0 & 0  & 0 & 0 & 0 & 0 & 0 & 0& 0 & 0 & 0 & 0 & 0 & 0 & 0  \\
1/2[631] &  &  & & 7.24 & 0 & 0 & $-0.22$ & 0 & 0 & 0 & 0.26 & 0 & 0 & 0 & 0 & 0.25 & 0 & 0 & 0 & 0 & 0 & 0& 0 & 0 & 0 & 0 & 0 & 0 & 0  \\
3/2[622] &  &  &  &  & 7.61 & 0 & 0 & $-0.18$ & 0 & 0 & 0 & 0 & 0 & 0 & 0 & 0 & 0.27 & 0   & 0 & 0 & 0   & 0& 0 & 0 & 0 & 0 & 0 & 0 & 0  \\

5/2[613]& & & & & & 8.01 & 0 & 0 & $-0.13$ & 0 & 0 & 0 & 0.20 & 0 & 0 & 0 & 0 & 0.25 & 0 & 0 & 0 & 0& 0 & 0 & 0 & 0 & 0 & 0 & 0  \\
1/2[620]& & & & & & & 7.56 & 0 & 0 & 0 & 0 & 0 & 0 & 0 & 0 & 0.29 & 0 & 0 & 0 & 0 & 0& 0& 0 & 0 & 0 & 0 & 0 & 0 & 0  \\
\hline
\end{tabular}

\vskip 0.3cm 

\begin{tabular}{ r r r r r r r r r r r r r r r r r r r r r r | r r r r r r r  }
                  &
${1\over 2}$[611] & ${1\over 2}$[600] & ${3\over 2}$[602] & ${1\over 2}$[631] & ${3\over 2}$[622] & ${5\over 2}$[613] & ${1\over 2}$[620] & 
${3\over 2}$[611] & ${5\over 2}$[602] & ${7\over 2}$[604] & ${1\over 2}$[651] & ${3\over 2}$[642] & ${5\over 2}$[633] & ${7\over 2}$[624] & 
${9\over 2}$[615] & ${1\over 2}$[640] & ${3\over 2}$[631] & ${5\over 2}$[622] & ${7\over 2}$[613] & ${9\over 2}$[604] & ${11\over 2}$[606]& 
${1\over 2}$[660] & ${3\over 2}$[651] & ${5\over 2}$[642] & ${7\over 2}$[633] & ${9\over 2}$[624] &${11\over 2}$[615] & ${13\over 2}$[606]\\

\hline

1/2[611]& 8.05 & $-0.16$ & 0 & 0.25 & 0 & 0 & 0.22 & 0 & 0 & 0 & 0 & 0  & 0  & 0 & 0 & 0  & 0 & 0 & 0 & 0 &  0 & 0 & 0 & 0 & 0 & 0 & 0 & 0 \\
1/2[600]& & 8.53 & 0 & 0 & 0  & 0  & 0.17  & 0  & 0  & 0  & 0  & 0  & 0  & 0 & 0 & 0  & 0 & 0 & 0 & 0 & 0 & 0& 0 & 0 & 0 & 0 & 0 & 0  \\
3/2[602] & & & 8.58 & 0 & 0  & 0  & 0  & 0.18  & 0  & 0  & 0  & 0 & 0  & 0 & 0 & 0  & 0 & 0 & 0 & 0 & 0 & 0& 0 & 0 & 0 & 0 & 0 & 0 \\
1/2[631] &  &  &  & 7.24 & 0  & 0  & $-0.22$ & 0 & 0 & 0 & 0.26 & 0 & 0 & 0 & 0 & 0.25 & 0 & 0 & 0 & 0 & 0 & 0& 0 & 0 & 0 & 0 & 0 & 0  \\
3/2[622] & & & & & 7.61 & 0 & 0 & $-0.18$ & 0 & 0 & 0 & 0 & 0 & 0 & 0 & 0 & 0.27 & 0 & 0 & 0 & 0 & 0& 0 & 0 & 0 & 0 & 0 & 0  \\

5/2[613] & & & & & & 8.01 & 0 & 0 & $-0.13$ & 0 & 0 & 0 & 0.20 & 0 & 0 & 0 & 0 & 0.25 & 0 & 0 & 0 & 0& 0 & 0 & 0 & 0 & 0 & 0  \\
1/2[620] & & & & & & & 7.56  & 0 & 0 & 0  & 0 & 0 & 0 & 0 & 0 & 0.29 & 0 & 0 & 0 & 0 & 0& 0& 0 & 0 & 0 & 0 & 0 & 0  \\
\hline
\end{tabular}

\end{table*}

\endgroup

\end{turnpage}



\begin{figure*}[htb]

\includegraphics[width=150mm]{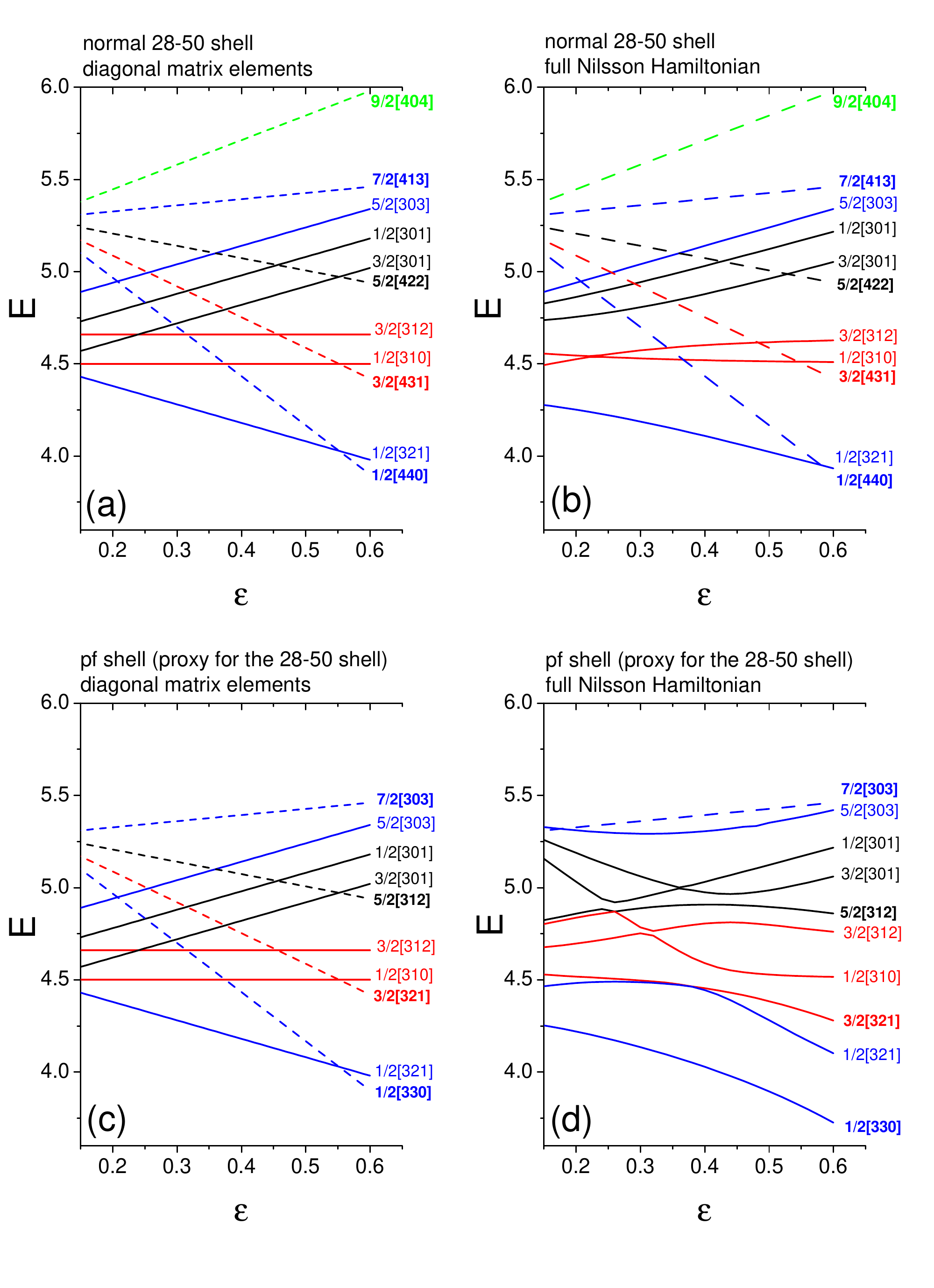}

\caption{(color online) Diagonal matrix elements (in units of $\hbar \omega_0$) of the Nilsson Hamiltonian for the 28-50 (a) and pf (c) proton or neutron shells compared 
to the results of the full diagonalization for the 28-50 (b) and pf (d) shells, as functions of the deformation parameter $\epsilon$. 
The Nilsson parameters are taken from Table I. The 1g$_{9/2}$ orbitals in (a) and (b), as well as the 1f$_{7/2}$ orbitals in (c), are 
indicated by dashed lines. The 1g$_{9/2}$ orbital labels in (a) and (b), as well as the 1f$_{7/2}$ orbital labels in (c) and (d), appear
in boldface. Orbitals are grouped in color only to facilitate visualizing 
the patterns of orbital evolution. 
 } 
\end{figure*}

\begin{figure*}[htb]

\includegraphics[width=150mm]{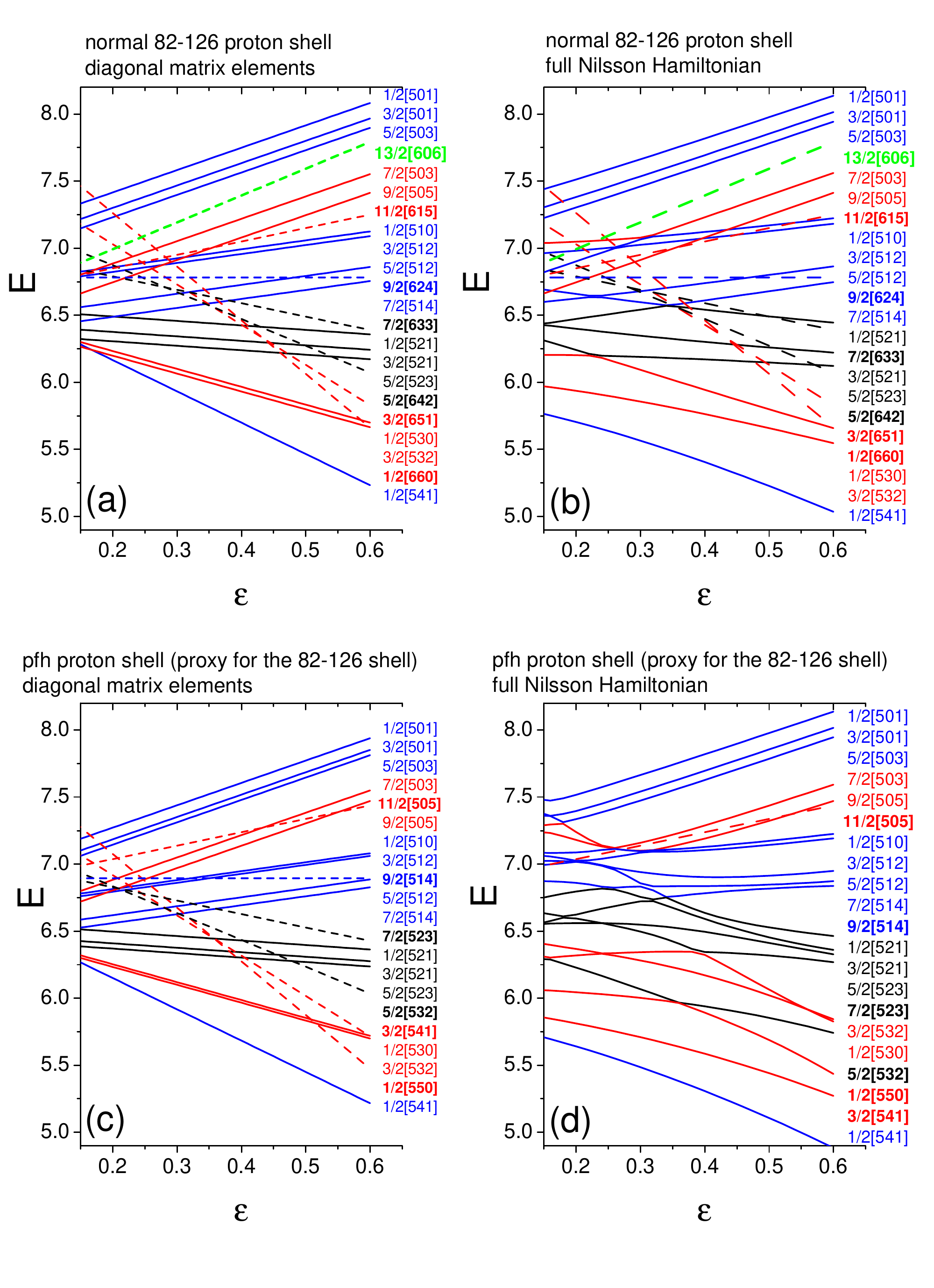}

\caption{(color online) Diagonal matrix elements (in units of $\hbar \omega_0$) of the Nilsson Hamiltonian for the 82-126 (a) and pfh (c) proton shells compared 
to the results of the full diagonalization for the 82-126 (b) and pfh (d) proton shells, as functions of the deformation parameter $\epsilon$. 
The Nilsson parameters are taken from Table I. The 1i$_{13/2}$ orbitals in (a) and (b), as well as the 1h$_{11/2}$ orbitals in (c), are 
indicated by dashed lines. The 1i$_{13/2}$ orbital labels in (a) and (b), as well as the 1h$_{11/2}$ orbital labels in (c) and (d), appear in boldface. Orbitals are grouped in color only to facilitate visualizing 
the patterns of orbital evolution. Note that the Nilsson labels at the right are always in the same order
as the energies of the orbitals as they appear at the right as well (largest deformation shown). 
Therefore, in some cases the order of the Nilsson orbitals changes slightly from panel to panel.
 } 
\end{figure*}

\begin{figure*}[htb]

\includegraphics[width=150mm]{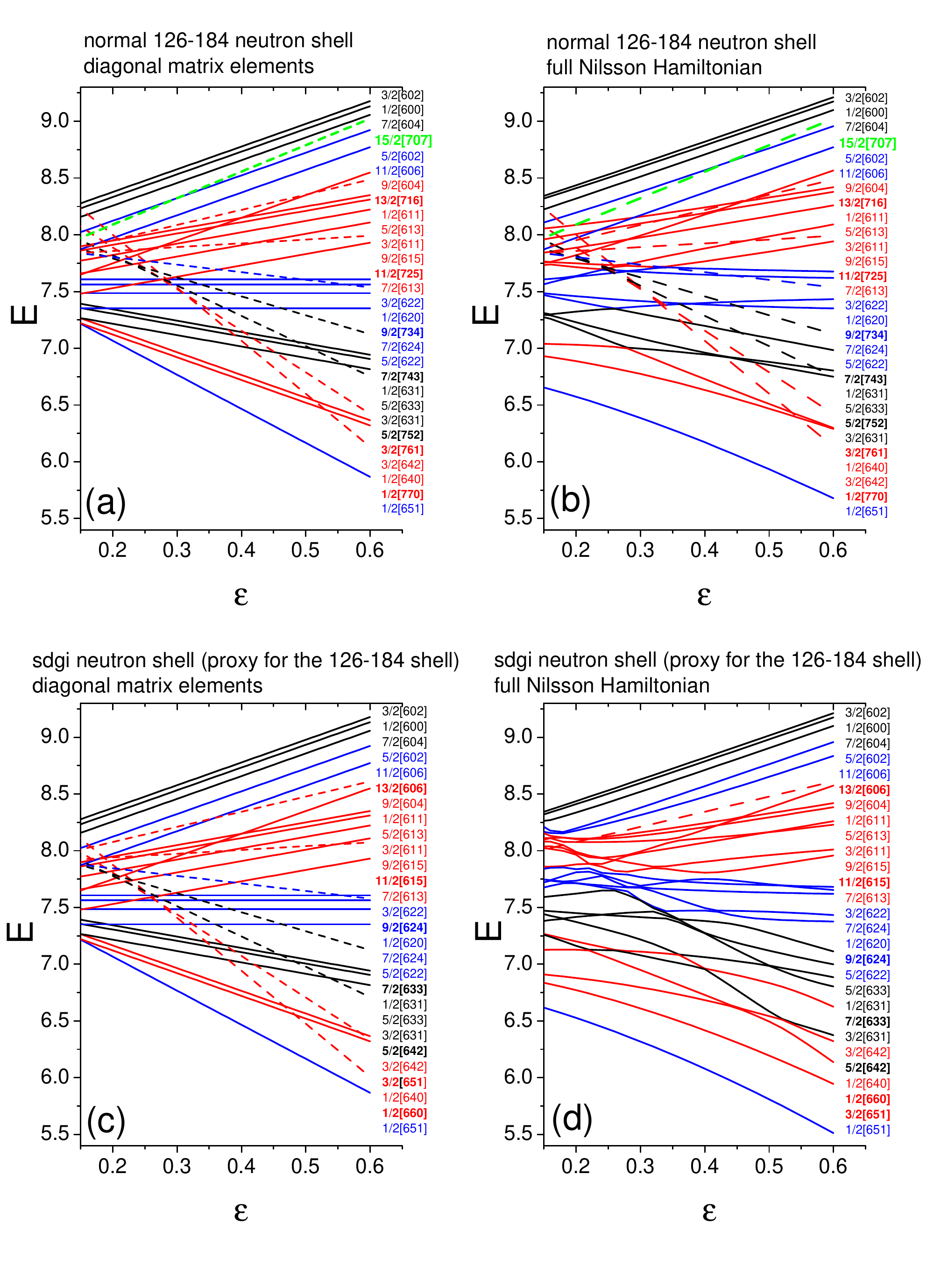}

\caption{(color online) Diagonal matrix elements (in units of $\hbar \omega_0$) of the Nilsson Hamiltonian for the 126-184 (a) and sdgi (c) neutron shells compared 
to the results of the full diagonalization for the 126-184 (b) and sdgi (d) neutron shells, as functions of the deformation parameter $\epsilon$. 
The Nilsson parameters are taken from Table I. The 1j$_{15/2}$ orbitals in (a) and (b), as well as the 1i$_{13/2}$ orbitals in (c), are 
indicated by dashed lines. The 1j$_{15/2}$ orbital labels in (a) and (b), as well as the 1i$_{13/2}$ orbital labels in (c) and (d), appear in boldface. Orbitals are grouped in color only to facilitate visualizing 
the patterns of orbital evolution. Note that the Nilsson labels at the right are always in the same order
as the energies of the orbitals as they appear at the right as well (largest deformation shown). 
Therefore, in some cases the order of the Nilsson orbitals changes slightly from panel to panel.
 } 
\end{figure*}

\section*{Acknowledgements} 
 
Support by the Bulgarian National Science Fund (BNSF) under Contract No. DFNI-E02/6 is gratefully acknowledged by N. M.
Work supported in part by the US DOE under grant number Grant No. DE-FG02- 91ER-40609, and by the FRIB laboratory. R.B.C. acknowledges support from the Max Planck Partner group, TUBA-GEBIP, and Istanbul University Scientific Research Project No. 26435.

\end{document}